\documentclass[
preprintnumbers, 
showpacs, 
amsmath, 
showkeys, 
amssymb, 
aps, 
superscriptaddress, 
twocolumn, 
longbibliography, 
letterpaper,
]{revtex4-2}
\usepackage{lipsum, babel}
\usepackage{color}
\usepackage{hyperref}
\usepackage{bm}
\usepackage{amsfonts}
\usepackage{mathrsfs}
\usepackage{booktabs}
\usepackage{graphicx}
\usepackage{amsmath}
\usepackage{float}
\usepackage{braket}
\usepackage{orcidlink}
\usepackage{natbib}

     \hypersetup{
    colorlinks=true,
    linkcolor=red,
    citecolor=blue,
}

\begin{document}

\newcommand{\sgn}{\operatorname{sgn}}
\newcommand{\hhat}[1]{\hat {\hat{#1}}}
\newcommand{\pslash}[1]{#1\llap{\sl/}}
\newcommand{\kslash}[1]{\rlap{\sl/}#1}
\newcommand{\lab}[1]{}
\newcommand{\sto}[1]{\begin{center} \textit{#1} \end{center}}
\newcommand{\rf}[1]{{\color{blue}[\textit{#1}]}}
\newcommand{\eml}[1]{#1}
\newcommand{\el}[1]{\label{#1}}
\newcommand{\er}[1]{Eq.\eqref{#1}}
\newcommand{\df}[1]{\textbf{#1}}
\newcommand{\mdf}[1]{\pmb{#1}}
\newcommand{\ft}[1]{\footnote{#1}}
\newcommand{\n}[1]{$#1$}
\newcommand{\fals}[1]{$^\times$ #1}
\newcommand{\new}{{\color{red}$^{NEW}$ }}
\newcommand{\ci}[1]{}
\newcommand{\de}[1]{{\color{green}\underline{#1}}}
\newcommand{\ke}{\rangle}
\newcommand{\br}{\langle}
\newcommand{\lb}{\left(}
\newcommand{\rb}{\right)}
\newcommand{\lbk}{\left[}
\newcommand{\rbk}{\right]}
\newcommand{\blb}{\Big(}
\newcommand{\brb}{\Big)}
\newcommand{\nn}{\nonumber \\}
\newcommand{\p}{\partial}
\newcommand{\pd}[1]{\frac {\partial} {\partial #1}}
\newcommand{\cd}{\nabla}
\newcommand{\cc}{$>$}
\newcommand{\bqa}{\begin{eqnarray}}
\newcommand{\eqa}{\end{eqnarray}}
\newcommand{\bqe}{\begin{equation}}
\newcommand{\eqe}{\end{equation}}
\newcommand{\bay}[1]{\left(\begin{array}{#1}}
\newcommand{\eay}{\end{array}\right)}
\newcommand{\eg}{\textit{e.g.} }
\newcommand{\ie}{\textit{i.e.}, }
\newcommand{\iv}[1]{{#1}^{-1}}
\newcommand{\st}[1]{|#1\ke}
\newcommand{\at}[1]{{\Big|}_{#1}}
\newcommand{\zt}[1]{\texttt{#1}}
\newcommand{\non}{\nonumber}
\newcommand{\m}{\mu}
\def\xa{{m}}
\def\xA{{m}}
\def\xb{{\beta}}
\def\xB{{\Beta}}
\def\xd{{\delta}}
\def\xD{{\Delta}}
\def\xe{{\epsilon}}
\def\xE{{\Epsilon}}
\def\xve{{\varepsilon}}
\def\xg{{\gamma}}
\def\xG{{\Gamma}}
\def\xk{{\kappa}}
\def\xK{{\Kappa}}
\def\xl{{\lambda}}
\def\xL{{\Lambda}}
\def\xo{{\omega}}
\def\xO{{\Omega}}
\def\xvp{{\varphi}}
\def\xs{{\sigma}}
\def\xS{{\Sigma}}
\def\xt{{\theta}}
\def\xvt{{\vartheta}}
\def\xT{{\Theta}}
\def \Tr {{\rm Tr}}
\def\CA{{\cal A}}
\def\CC{{\cal C}}
\def\CD{{\cal D}}
\def\CE{{\cal E}}
\def\CF{{\cal F}}
\def\CH{{\cal H}}
\def\CJ{{\cal J}}
\def\CK{{\cal K}}
\def\CL{{\cal L}}
\def\CM{{\cal M}}
\def\CN{{\cal N}}
\def\CO{{\cal O}}
\def\CP{{\cal P}}
\def\CQ{{\cal Q}}
\def\CR{{\cal R}}
\def\CS{{\cal S}}
\def\CT{{\cal T}}
\def\CV{{\cal V}}
\def\CW{{\cal W}}
\def\CY{{\cal Y}}
\def\BC{\mathbb{C}}
\def\BR{\mathbb{R}}
\def\BZ{\mathbb{Z}}
\def\sA{\mathscr{A}}
\def\sB{\mathscr{B}}
\def\sF{\mathscr{F}}
\def\sG{\mathscr{G}}
\def\sH{\mathscr{H}}
\def\sJ{\mathscr{J}}
\def\sL{\mathscr{L}}
\def\sM{\mathscr{M}}
\def\sN{\mathscr{N}}
\def\sO{\mathscr{O}}
\def\sP{\mathscr{P}}
\def\sR{\mathscr{R}}
\def\sQ{\mathscr{Q}}
\def\sS{\mathscr{S}}
\def\sX{\mathscr{X}}


\title{High $Q^2$  Behavior of the Proton Structure Function through the Balitsky-Kovchegov Equation}
\author{Wei Kou\orcidlink{0000-0002-4152-2150}}
\email{kouwei@impcas.ac.cn}
\affiliation{Institute of Modern Physics, Chinese Academy of Sciences, Lanzhou 730000, China}
\affiliation{School of Nuclear Science and Technology, University of Chinese Academy of Sciences, Beijing 100049, China}
\author{Gang Xie}
\affiliation{Institute of Modern Physics, Chinese Academy of Sciences, Lanzhou 730000, China}
\affiliation{University of Chinese Academy of Sciences, Beijing 100049, China}
\author{Xiaopeng Wang}
\affiliation{Institute of Modern Physics, Chinese Academy of Sciences, Lanzhou 730000, China}
\affiliation{School of Nuclear Science and Technology, University of Chinese Academy of Sciences, Beijing 100049, China}
\affiliation{Lanzhou University, Lanzhou 730000, China}
\author{Chengdong Han}
\email{chdhan@impcas.ac.cn (corresponding\,author)}
\affiliation{Institute of Modern Physics, Chinese Academy of Sciences, Lanzhou 730000, China}
\affiliation{School of Nuclear Science and Technology, University of Chinese Academy of Sciences, Beijing 100049, China}
\author{Xurong Chen}
\email{xchen@impcas.ac.cn (corresponding\,author)}
\affiliation{Institute of Modern Physics, Chinese Academy of Sciences, Lanzhou 730000, China}
\affiliation{University of Chinese Academy of Sciences, Beijing 100049, China}
\affiliation{Southern Center for Nuclear Science Theory (SCNT), Institute of Modern Physics, Chinese Academy of Sciences, Huizhou 516000, Guangdong Province, China}


\begin{abstract}

Numerous experimental and theoretical investigations have highlighted the power law behavior of the proton structure function $ F_2(x, Q^2) $, particularly the dependence of its power constant on various kinematic variables. In this study, we analyze the proton structure function $ F_2 $ employing the analytical solution of the Balitsky-Kovchegov equation, with a focus on the high $ Q^2 $ regime and small $ x $ domains. Our results indicate that as $ Q^2 $ increases, the slope parameter $ \lambda $, which characterizes the growth rate of $ F_2 $, exhibits a gradual decrease, approaching a limiting value of $ \lambda \approx 0.41 \pm 0.01 $ for large $ Q^2 $. We suggest that this behavior of $ \lambda $ may be attributed to mechanisms such as gluon overlap and the suppression of phase space growth. To substantiate these conclusions, further high-precision electron-ion collision experiments are required, encompassing a broad range of $ Q^2 $ and $ x $.

\end{abstract}

                        
 \keywords{Deep Inelastic Scattering, Structure Functions, Balitsky-Kovchegov Equation, QCD}
\maketitle


\section{Introduction}
\label{sec:intro}


Over the last two decades, the leading order (LO) and next-to-leading order (NLO) formulation of the Balitsky-Fadin-Kuraev-Lipatov (BFKL) equation \cite{Lipatov:1976zz,Kuraev:1977fs,Balitsky:1978ic} have been examined by L. N. Lipatov et al. \cite{Ellis:2008yp,Kowalski:2010ue,Kowalski:2011zza,Kowalski:2012ur,Kowalski:2014iqa,Kowalski:2015paa,Kowalski:2017umu,Salam:1998tj} using the Green's function approach. They contended that the BFKL equation should be regarded as an eigen-equation with an eigenvalue $ \omega_n $. The authors in Ref. \cite{Ellis:2008yp} argued that the discrete, asymptotically-free BFKL Pomeron was initially shown to describe HERA data at low $ x $ and high $ Q^2 $. Most of the deep inelastic scattering (DIS) data provided by HERA are well described by the Dokshitzer–Gribov–Lipatov–Altarelli–Parisi (DGLAP) equations \cite{Dokshitzer:1977sg,Gribov:1972ri,Lipatov:1974qm,Altarelli:1977zs}, which are presented through the lens of renormalization group evolution \cite{Wilson:1974mb}. Nonetheless, the discussion surrounding HERA data in the small $ x $ region has initiated the combination of the DGLAP and BFKL equations. It should be noted that the variation of the scattering cross-section with energy must satisfy the unitarity limit naturally; conversely, the BFKL equation, as a linear evolution equation, could challenge this property in the high-energy region. Fortunately, the infinite growth of gluons appears to be suppressed by certain processes. The saturation behavior of gluons is discussed in Refs. \cite{Mueller:2001fv,Stasto:2000er,Munier:2003vc} (and references cited therein). The Balitsky-Kovchegov (BK) equation \cite{Balitsky:1997mk,Kovchegov:1999yj,Kovchegov:1999ua,Balitsky:2001re} represents a nonlinear evolution of gluons and serves as a mean-field approximation of the Jalilian-Marian-Iancu-McLerran-Weigert-Leonidov-Kovner (JIMWLK) equation \cite{Balitsky:1995ub,Jalilian-Marian:1997qno,Iancu:2000hn,Weigert:2000gi}. The BK equation effectively describes gluon behavior in the small $ x $ region. In a previous work, some of us obtained an analytical solution of the BK equation in momentum space through a novel approach \cite{Wang:2020stj}. 

The HERA collaboration provides high-precision data for the proton structure function $ F_2 $, making it a key resource for studying the behavior of $ F_2 $ under small $ x $ and large $ Q^2 $ conditions. The behavior of the structure function in the small $ x $ region is generally understood by analyzing its slope parameter $ \lambda $ \cite{Bartels:2000ze}, defined as $ \lambda \propto \partial \ln F_{2} / \partial \ln (1 / x) $. A significant body of previous work comprehensively addresses this topic \cite{Ellis:2008yp,Kowalski:2010ue,Kowalski:2011zza,Kowalski:2012ur,Kowalski:2014iqa,Kowalski:2015paa,Kowalski:2017umu,Salam:1998tj,Cvetic:2009kw,Cooper-Sarkar:1997pqx,Kotikov:1998qt,Kotikov:2007ua,Ball:1994kc,Illarionov:2004nw,Mankiewicz:1996sd} (and references cited therein). In recent years, physicists have offered various explanations for and fitting results related to the new HERA data \cite{H1:2015ubc,Luszczak:2019dsp}. Compared to the range of $ Q^2 $ explored in the data, several theoretical studies have provided results for the parameter $ \lambda $ as a function of $ Q^2 $ only in the intermediate region, such as $ \mathcal{O}(200) \text{ MeV}^2 $ \cite{Kotikov:1998qt,Illarionov:2004nw,Kaidalov:2000wg,Donnachie:2003cs}. It is important to note that the structure function $ F_2 $ data at very high $ Q^2 $ pertain predominantly to the large $ x $ scale. These considerations lead us to explore the double-asymptotic-scaling (DAS) phenomenon, which describes the limits as $ Q^2 \to \infty $ and $ x \to 0 $. 

In the present work, we combine experimental findings and QCD evolution theory to analyze the proton structure function from the perspective of the BK equation. Based on the analytical solution presented in previous work \cite{Wang:2020stj}, we calculate the proton structure function $ F_2 $ at higher $ Q^2 $ and analyze the small $ x $ behavior by fitting the computed results. The organization of this paper is as follows: In Sec. \ref{sec:theory}, we briefly introduce the methods and principles for calculating the proton structure function $ F_2 $ using the analytical solution of the BK equation from Ref. \cite{Wang:2020stj}. In Sec. \ref{sec:F2}, we discuss how to obtain the structure function $ F_2 $ and its corresponding effective slope parameter $ \lambda $ at high energies. In Sec. \ref{sec:dissc}, we highlight the interesting results from Sec. \ref{sec:F2}. Finally, the conclusions of this work are presented in Sec. \ref{sec:conclu}.

\section{Formalism}
\label{sec:theory}
\begin{figure}[htbp]
	\centering
	\includegraphics[width=0.45\textwidth]{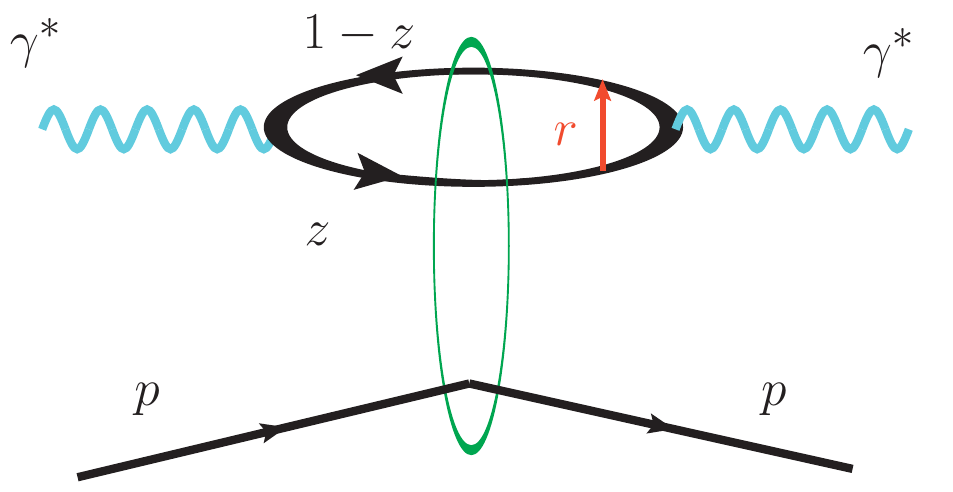}
	\caption{(color online).
		The elastic scattering of a virtual photon
		on a proton in the dipole representation \cite{Kowalski:2006hc}.}
	\label{fig:diff-xsection-proton-target}
\end{figure}

We consider the high-energy scattering process using the dipole model (see FIG. \ref{fig:diff-xsection-proton-target}). The high-speed moving photon fluctuates a $q\bar{q}$ pair from the QCD vacuum, forming a dipole. The total cross section for the overall $\gamma^*p$ scattering is represented by integrating the dipole cross section with the photon wave functions \cite{Mueller:1994jq,Kowalski:2003hm,Kowalski:2006hc}. The momentum space formulation of the dipole scattering cross section can be obtained through Fourier transform. If the transverse profile of the proton is considered as an isotropic disk with radius $R_p$, the dipole scattering cross-section is proportional to the forward scattering amplitude $\mathcal{N}(r,Y)$, which depends on the dipole transverse size $r$ and rapidity $Y$, as described in \cite{deSantanaAmaral:2006fe}:

\begin{equation}
    \sigma_{\mathrm{dip}}^{\gamma^{*} p}(r, Y) = 2 \pi R_{p}^{2} \mathcal{N}(r, Y),
    \label{eq:disk-rp}
\end{equation}
where $ R_p $ is taken to be the electromagnetic radius of the proton.

Based on the above information, the structure function of proton can be expressed in momentum space as \cite{Barone:2002cv,deSantanaAmaral:2006fe}
\begin{equation}
	\small
	F_{2}\left(x, Q^{2}\right)=\frac{Q^{2} R_{p}^{2} N_{c}}{4 \pi^{2}} \int_{0}^{\infty} \frac{d k}{k} \int_{0}^{1} d z\left|\tilde{\Psi}\left(k^{2}, z ; Q^{2}\right)\right|^{2} \mathcal{N}(k, Y),
	\label{eq:F2-BK}
\end{equation}
with the Fourier transform
\begin{equation}
	\small
	\mathcal{N}(k, Y)=\frac{1}{2 \pi} \int \frac{d^{2} r}{r^{2}} e^{i \mathbf{k} \cdot \mathbf{r}} \mathcal{N}(r, Y)=\int_{0}^{\infty} \frac{d r}{r} J_{0}(k r)\mathcal{N}(r, Y).
	\label{eq:fourier}
\end{equation}
The photon wave functions $ \tilde{\Psi} $ in Eq. (\ref{eq:F2-BK}) can be found in Ref. \cite{deSantanaAmaral:2006fe} in momentum space. It is important to note that not only DIS calculations but also the vector meson diffractive production process has a similar wave function representation for calculations \cite{Kowalski:2006hc,Xie2015}. 

We will now handle the forward scattering amplitude using the method described in \cite{Wang:2020stj}. The BK equation is rewritten as the Fisher-KPP equation \cite{https://doi.org/10.1111/j.1469-1809.1937.tb02153.x} with appropriate variable substitutions. More specifically, the BK equation in momentum space can be described as a nonlinear variant of the BFKL equation \cite{Munier:2003vc},
\begin{equation}
	\frac{\partial\mathcal{N}(k,Y)}{\partial Y}=\frac{\alpha_\mathrm{s}N_\mathrm{c}}{\pi}\chi\left(-\frac{\partial}{\partial\ln k^2}\right)\mathcal{N}(k,Y)-\frac{\alpha_\mathrm{s}N_\mathrm{c}}{\pi}\mathcal{N}^2(k,Y),
	\label{eq:BK}
\end{equation}
where $\chi(\lambda)=\psi(1)-\frac{1}{2}\psi\left(1-\frac{\lambda}{2}\right)-\frac{1}{2}\psi\left(\frac{\lambda}{2}\right)$ is the BFKL kernel with the digamma function $\psi(\lambda)=\Gamma^{\prime}({\lambda})/\Gamma(\lambda)$. In addition, the $\alpha_\mathrm{s}$ is the strong coupling. In this work, we choose $Y=\bar{\alpha_{\mathrm{s}}}\log(1/x)$ with $\bar{\alpha_{\mathrm{s}}}=\alpha_{\mathrm{s}}N_c/\pi$.

We utilize the analytical solution of the BK equation obtained through the Homogeneous Balance method in Refs. \cite{WANG1995169,WANG1996279,ZHOU200331,ZHOU200477,wang2014simplified}, which can be expressed as \cite{Wang:2020stj}
\begin{equation}
    \mathcal{N}(L, Y) = \frac{A_{0} e^{5 A_{0} Y / 3}}{\left[e^{5 A_{0} Y / 6}+e^{\left[-\theta+\sqrt{A_{0} / 6 A_{2}}\left(L-A_{1} Y\right)\right]}\right]^{2}}.
    \label{eq:BK-solution}
\end{equation} 
Here, the variable $ L = \ln \left(k^{2} / k_{0}^{2}\right) $ with the cutoff $ k_0 = \Lambda_{\mathrm{QCD}} = 200 $ MeV. The parameter values $ \theta $, $ A_0 $, $ A_1 $, and $ A_2 $ are referenced in \cite{Marquet:2005zf,Wang:2020stj} and can be obtained by fitting experimental data.

\section{Structure function $\mathbf{F_2(x,Q^2)}$ in High energies}
\label{sec:F2}
In Ref. \cite{deSantanaAmaral:2006fe}, the authors discussed in detail the process of calculating the structure function $ F_2 $, including the selection of parameters contained in the momentum space photon wave function. We employ the same formalism from Ref. \cite{deSantanaAmaral:2006fe} to derive several parameters in the analytical solution of the BK equation by fitting HERA data \cite{H1:2013ktq,H1:2015ubc}. Considering the small $ x $ limit, we chose to fit the experimental data in the range $ 2.5 \, \text{GeV}^2 < Q^2 < 250 \, \text{GeV}^2 $ because the quantity of data and the precision in this region are suitable.

\begin{figure*}[htbp]
	\centering
	\includegraphics[width=0.96\textwidth]{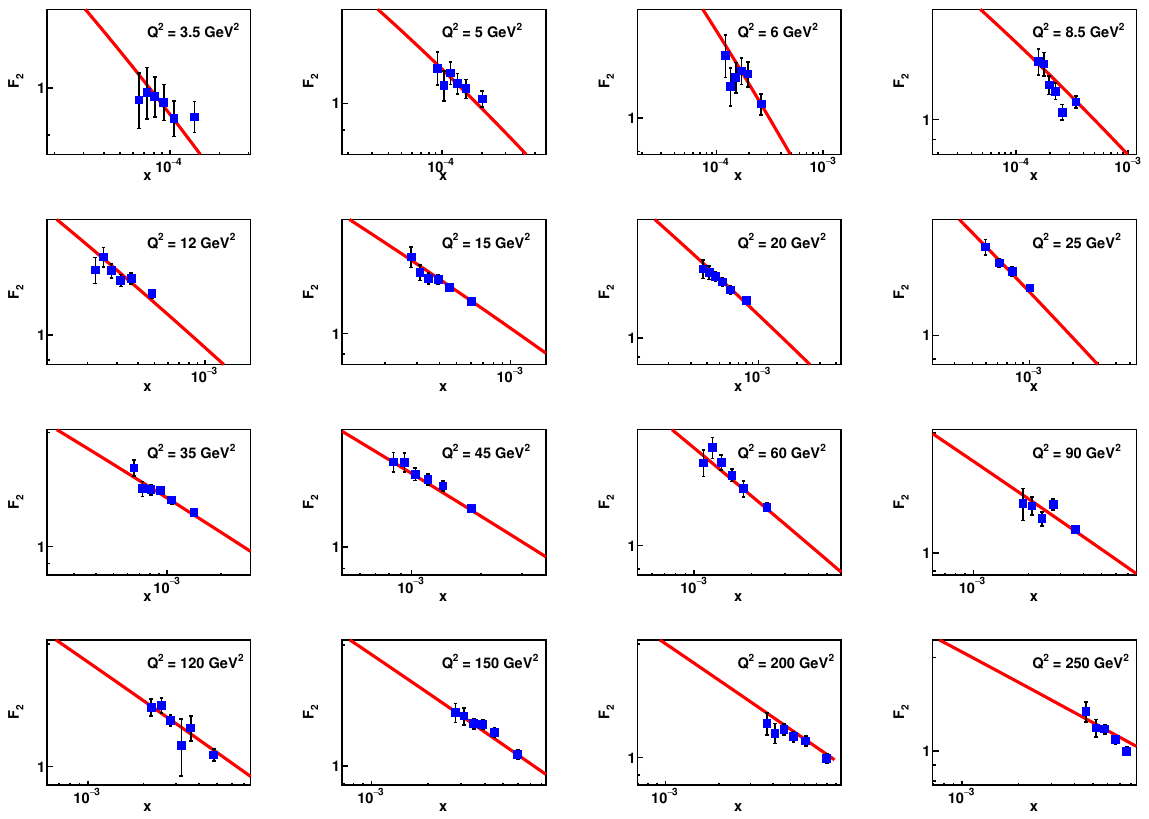}
	\caption{(color online). $x$ dependence of structure function $F_2(x,Q^2)$ with bins of $Q^2$. The data from H1 \cite{H1:2013ktq} (blue squares) are compared with our fit by analytical solution of BK \cite{Wang:2020stj} (red lines). The parameters determination is discussed in text.}
	\label{fig:fit}
\end{figure*}

According to the discussions in \cite{Wang:2020stj}, we fixed the strong coupling constant $\bar{\alpha_{\mathrm{s}}} = 0.191$, which is effective in the energy range we considered. We need to emphasize that the strong coupling constant $\bar{\alpha_{\mathrm{s}}}$ is fixed because the primary focus of this work is on the behavior of the structure function at higher $Q^2$, where the coupling constant can be considered to have a negligible impact on the results. This simplification is well justified. The chosen value for the coupling constant has also been used to calculate the diffractive production processes of vector mesons \cite{Wang:2022jwh}, successfully describing experimental data \cite{ZEUS:2004yeh,ZEUS:2007iet,H1:1999pji,H1:2005dtp,H1:2009cml}. 
To calculate $F_2$ using Eq. (\ref{eq:F2-BK}), the color number $N_c$ is set to 3, and the considerations regarding quark mass and flavor remain unchanged from those we used in Ref. \cite{deSantanaAmaral:2006fe}. Specifically, we set $m_u = m_d = m_q = 140\,\text{MeV}$ and $m_c = 1.4\,\text{GeV}$ for the quark masses \cite{deSantanaAmaral:2006fe}, while the proton size is treated as the electromagnetic radius $R_p = 4.23\,\text{GeV}^{-1}$. For the selection of quark mass parameters in the photon wave function, we referred to the work of the GBW model \cite{Golec-Biernat:1998zce,Bartels:2002cj,Kowalski:2003hm}. 
In principle, the choice of these parameters can influence the results; however, we treat the photon wave function as a priori and base our discussion of the applicability of the BK equation’s analytical solution on this assumption. Based on these fixed parameter selections, we utilize the HERA data to fit and obtain the parameters in Eq. (\ref{eq:BK-solution}). The fitting results are displayed in FIG. \ref{fig:fit} (with partial fitting results), and the fitting parameters of Eq. (\ref{eq:BK-solution}) are determined as $A_0 = 0.5262 \pm 0.0024$, $A_1 = 1.4383 \pm 0.0133$, $A_2 = 0.1047 \pm 0.0008$, and $\theta = -0.4572 \pm 0.0110$.

Due to various experimental constraints, researchers are often particularly interested in the physics at the energy limits, such as $ x \to 0 $ and $ Q^2 \to \infty $. Although this research may appear more mathematical in nature, it frequently yields intriguing predictions. In this study, we aim to investigate the small-$ x $ behavior of $ F_2 $ at higher $ Q^2 $. 
The slope parameter $ \lambda $ can be expressed as $ \lambda \propto \partial \ln F_{2} / \partial \ln (1/x) $ at different $ Q^2 $ values, as discussed in Ref. \cite{Cvetic:2009kw}. In the small-$ x $ range ($ x < 0.01 $), the structure function of the proton $ F_2 $ can be parameterized in a power-like form:
\begin{equation}
    F_{2}\left(x, Q^{2}\right) = C x^{-\lambda\left(Q^{2}\right)}.
    \label{eq:F2-power}
\end{equation}
According to our calculations, the slope parameter $ \lambda $ is determined from the computed $ F_2(x, Q^2) $ values using the BK equation and the dipole amplitude, in accordance with Eqs. (\ref{eq:F2-BK}) and (\ref{eq:BK-solution}). This will be discussed in the next section.

\section{Results and discussions}
\label{sec:dissc}
\begin{figure}[htbp]
	\centering
	\includegraphics[width=0.48\textwidth]{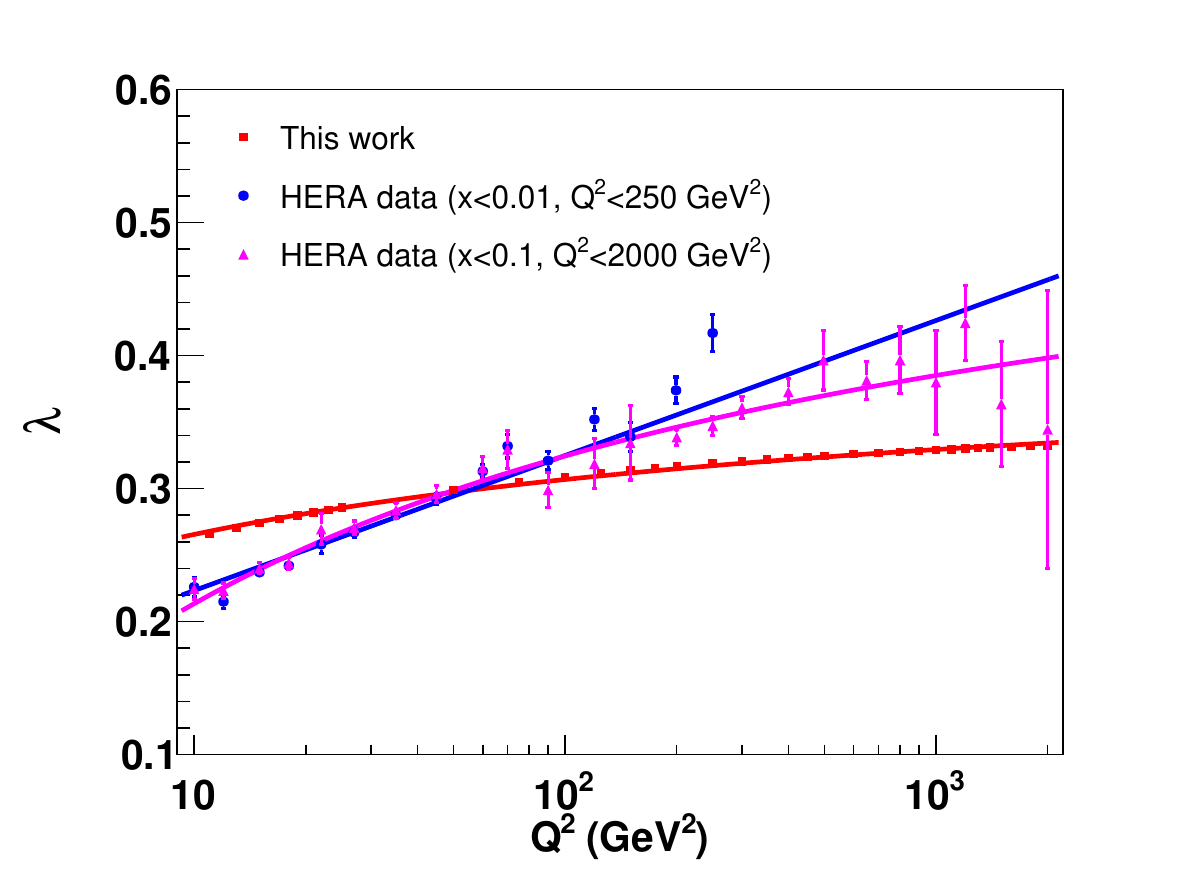}
	\caption{(color online). Slope parameter $\lambda$ v.s. energy scale $Q^2$. The  red squares with error bar represent our calculations, determined by Eqs. (\ref{eq:F2-BK}), (\ref{eq:BK-solution}) and (\ref{eq:F2-power}). The Blue circles denote the extracted $\lambda$ from \cite{H1:2015ubc,Luszczak:2019dsp} ($x<0.01$, $Q^2<250$ GeV$^2$) and magenta triangles are extracted from same data from HREA \cite{H1:2015ubc} ($x<0.1$, $Q^2<2000$ GeV$^2$). Red line represents the fitting result comes from Eq. (\ref{eq:lambda-fit1}) which describe the red points. The blue and magenta lines are fitting results of the corresponding color points respectively.}
	\label{fig:lambda-Q2}
\end{figure}

The BK equation, as a QCD theory, possesses predictive power under limiting conditions. Previous studies have generated theoretical predictions based on experimental data and appropriate assumptions, which have proven to be both effective and inspiring. In this work, we take this a step further by extending both the Bjorken scale $ x $ and $ Q^2 $ to their limits, utilizing the analytical solution of the BK equation.
As we have discussed, we aim to characterize the behavior of $ F_2 $ under very high $ Q^2 $ conditions. We can readily extend $ F_2 $ to $ Q^2 \sim \mathcal{O}(2000) \, \text{GeV}^2 $ through numerical calculations within the range $ 2.0 \times 10^{-5} < x < 1.0 \times 10^{-2} $. Utilizing the analytical solution (\ref{eq:BK-solution}) with the fitted parameters, the behavior of the slope factor $ \lambda $ is illustrated in FIG. \ref{fig:lambda-Q2}.

Figure \ref{fig:lambda-Q2} presents the $ \lambda $ values extracted from the HERA data (blue and magenta points) \cite{H1:2013ktq,H1:2015ubc} alongside our calculated results using the BK equation (red solid squares). We also provide parameterized formulas to describe the $ \lambda - Q^2 $ relation, expressed as:
\begin{equation}
    \lambda(Q^2) = a\left(1 - \frac{b}{\ln(Q^2/\Lambda_{\mathrm{QCD}}^2)}\right) + c,
    \label{eq:lambda-fit1}
\end{equation}
for the red and magenta points in FIG. \ref{fig:lambda-Q2}, and 
\begin{equation}
    \lambda(Q^2) = A\ln\left(\frac{Q^2}{\Lambda_{\mathrm{QCD}}^2}\right) + B,
    \label{eq:lambda-fit2}
\end{equation}
for the blue points \cite{Luszczak:2019dsp}. In these equations, the QCD cutoff $ \Lambda_{\mathrm{QCD}} = 200 $ MeV is fixed. We use Eq. (\ref{eq:lambda-fit1}) to fit our calculations and the HERA data \cite{H1:2015ubc} (magenta triangles, $ Q^2 < 2000 \, \text{GeV}^2 $). All parameter information is provided in TABLE \ref{tab:para}, TABLE \ref{tab:data1}, and TABLE \ref{tab:data2}, respectively.
    \begin{table*}[htbp]
	\renewcommand\arraystretch{1}
	\setlength\tabcolsep{5pt}
	\begin{center}
		\caption{Parameters determination according to Eq. (\ref{eq:lambda-fit1}) and (\ref{eq:lambda-fit2}), using our calculations and HERA data. The results of the last two lines come from Refs. \cite{H1:2015ubc,Luszczak:2019dsp}.}
        \label{tab:para}
		\begin{tabular}{ |c|c|c|c|c|c|c| }
			\hline
			Parameters & $\Lambda_{\mathrm{QCD}}$ (MeV)& $a$  & $b$ & $c$ & $A$ & $B$ \\
			\hline
			This work&200 & $0.162\pm0.006$ &$4.800\pm0.168$  & $0.244\pm0.006$& &\\
			\hline
			Fits from \cite{Luszczak:2019dsp}&200 &  & & &$0.044\pm0.001$ &$-0.020\pm0.004$\\
			\hline
			Fits from \cite{H1:2015ubc}&200 & $0.195\pm0.017$ &$10.657\pm0.838$ &$0.395\pm0.016$ & &\\
			\hline
		\end{tabular}	
	\end{center}
\end{table*}

\begin{table}[htbp]
	\begin{center}
		\renewcommand\arraystretch{1}
		\setlength\tabcolsep{25pt}
			\caption{The results for the $\lambda$ constant obtained from the fits of the function (\ref{eq:F2-power}) from Ref. \cite{Luszczak:2019dsp} ($x<0.01$), $Q^2<250$ GeV$^2$.}
            		\label{tab:data1}
		\begin{tabular}{ |c|c|c|}
			\hline
			$Q^2$ (GeV$^2$) & $\lambda$ & $\delta \lambda$\\
			\hline
			0.35&	0.110	&	0.008\\
			\hline
			0.4&	0.082&		0.009\\
			\hline
			0.5&	0.100	&	0.009\\
			\hline
			0.65	&0.121	&	0.011\\
			\hline
			0.85&	0.150	&	0.014\\
			\hline
			1.2&	0.133	&	0.013\\
			\hline
			1.5&	0.142	&	0.009\\
			\hline
			2&	0.159	&	0.007\\
			\hline
			2.7&	0.169	&	0.005\\
			\hline			 
			3.5&	0.173	&	0.004\\
			\hline
			4.5&	0.189	&	0.004\\
			\hline
			6.5&	0.200	&	0.003\\
			\hline
			8.5	&0.208	&	0.004\\
			\hline
			10&0.226	&	0.007\\
			\hline
			12	&0.215	&	0.005\\
			\hline
			15	&0.237	&	0.003\\
			\hline
			18	&0.242	&	0.003\\
			\hline
			22	&0.258	&	0.007\\
			\hline
			27	&0.267	&	0.004\\
			\hline
			35	&0.280	&	0.003\\
			\hline
			45	&0.292	&	0.004\\
			\hline
			60	&0.313	&	0.005\\
			\hline
			70	&0.332	&	0.009\\
			\hline
			90	&0.321	&	0.007\\
			\hline
			120&	0.352&		0.008\\
			\hline
			150&	0.339&		0.011\\
			\hline
			200&	0.373&		0.010\\
			\hline
			250&	0.417&		0.014 \\
			\hline
		\end{tabular}
	\end{center}
\end{table}
\begin{table}[htbp]
	\begin{center}
		\renewcommand\arraystretch{1}
		\setlength\tabcolsep{25pt}
				\caption{The results for the $\lambda$ constant obtained from the fits of the function (\ref{eq:F2-power}) from Ref. \cite{H1:2015ubc} ($x<0.1$, $Q^2<2000$ GeV$^2$).}
                		\label{tab:data2}
		\begin{tabular}{ |c|c|c|}
			\hline
			$Q^2$ (GeV$^2$) & $\lambda$ & $\delta \lambda$\\
			\hline
			10 &0.224 & 0.008\\
			\hline
			12& 0.223 & 0.006\\
			\hline
			15 &0.240 & 0.004\\
			\hline
			18 &0.244&  0.005\\
			\hline
			22 &0.269&  0.012\\
			\hline
			27 &0.271&  0.005\\
			\hline
			35 &0.284& 0.006\\
			\hline
			45& 0.296&  0.007\\
			\hline
			60 &0.316 & 0.008\\
			\hline
			70 &0.330&  0.014\\
			\hline
			90& 0.299 & 0.013\\
			\hline
			120 &0.319&  0.019\\
			\hline		
			150 &0.334 & 0.0281\\
			\hline
			200& 0.339 & 0.006\\
			\hline
			250& 0.347 & 0.007\\
			\hline		
			300 &0.361 & 0.008\\
			\hline	
			400&0.373 & 0.010\\
			\hline
			500& 0.396 & 0.023\\
			\hline
			650 &0.381 & 0.014\\
			\hline
			800& 0.397 & 0.025\\
			\hline
			1000 &0.380 &0.039\\
			\hline
			1200 &0.425&  0.028\\
			\hline
			1500 &0.364&  0.047\\
			\hline
			2000 &0.345&  0.105\\
			\hline
		\end{tabular}
	\end{center}
\end{table}

 Let us consider this formula. In contrast to the logarithmic growth of $ \lambda $ with $ Q^2 $ discussed in other works \cite{Kaidalov:2000wg,Donnachie:2003cs}, we find that the rate of increase of $ \lambda $ decreases slowly with increasing $ Q^2 $, as indicated by $ \mathrm{d}\lambda/\mathrm{d}Q^2\big|_{Q^2\to \infty} = 0 $. This behavior resembles that of the running strong coupling constant in the extreme high-energy limit (see Ref. \cite{Schrempp:2005vc} and references cited therein). 
From Eq. (\ref{eq:lambda-fit1}), we consider the scenario where $ Q^2 $ approaches infinity. At this point, $ \lambda $ shows minimal growth with respect to $ Q^2 $, and its value reaches $ \lambda(Q^2\to \infty) = a + c \simeq 0.41 \pm 0.01 $. This value may represent the upper limit that $ \lambda $ will attain. The ``frozen" $ \lambda $ signifies that there is no change even if $ Q^2 $ continues to rise, leading to fixed rates of variation for the $ x $-dependence of $ F_2 $. We also examined the selection of the fitting range for $ F_2 $ data and found no significant differences in the parameters $ C $ and $ \lambda $ (as defined in Eq. (\ref{eq:F2-power})) between $ x < 0.01 $ and $ x < 0.001 $ under these conditions.

To explain the ``frozen" $ \lambda $, we investigate the behavior of gluons in high-energy $ \gamma^* p $ scattering using the BK equation. The parameter $ \lambda $ serves as a power index that represents the rate of gluon emissions per unit of rapidity from a high-speed dipole \cite{Bartels:2000ze,Luszczak:2019dsp}. The formation of the structure function is expressed as \cite{Ellis:2008yp,Kowalski:2010ue}:
\begin{equation}
    F_{2}\left(x, Q^{2}\right) = \int_{x}^{1} dz \int \frac{dk}{k} \Phi_{\text{DIS}}(z, Q, k) \, x g\left(\frac{x}{z}, k\right),
    \label{eq:gluon distribution-F2}
\end{equation}
where $ x g\left(\frac{x}{z}, k\right) $ denotes the unintegrated gluon density defined in \cite{Ellis:2008yp}, expressed in the form of a power law with index $ \lambda $ (or the eigenvalue of the BFKL equation $ \omega_n $). On one hand, the invariant $ \lambda $ in the high $ Q^2 $ region indicates that the gluon emission rate has reached equilibrium. The behavior of $ \lambda $ we obtained aligns well with discussions and results from previous works \cite{Martin:1996as,Golec-Biernat:1998zce,Golec-Biernat:1999qor}. On the other hand, as $ Q^2 $ increases and $ x $ decreases \cite{Luszczak:2019dsp}, the phase space for gluon emissions expands rapidly. When the number of gluons becomes large, they begin to overlap, which limits the growth of phase space and ultimately results in a particular value, consistent with the requirements of unitarity in QCD theory.

Our results are built upon previous related works \cite{Ellis:2008yp,Kowalski:2010ue,Kowalski:2011zza,Kowalski:2012ur,Kowalski:2014iqa,Kowalski:2015paa,Kowalski:2017umu,Salam:1998tj,Cvetic:2009kw,Cooper-Sarkar:1997pqx,Kotikov:1998qt,Kotikov:2007ua,Ball:1994kc,Illarionov:2004nw,Mankiewicz:1996sd}, which generalize the $ Q^2 $-dependence of $ \lambda $ to higher $ Q^2 $ situations. However, in this work, we have made certain approximations, such as the selection of fixed strong coupling constants. This approximation implies that we consider the calculation results to be more significant at specific $ Q^2 $ values, while at very low or high $ Q^2 $, discrepancies may arise (see FIG. \ref{fig:lambda-Q2}).
We argue that it is feasible to use the BK equation to investigate the proton structure function, despite these approximations. Our results are grounded in QCD evolution theory, and the BK equation has been proven to be a valuable tool in the study of  DIS and diffraction processes \cite{Mueller:2001fv,Stasto:2000er,Munier:2003vc}. The challenges we encountered are expected to be addressed in future research, particularly through the application of the BK equation with running coupling constants \cite{Marquet:2005ic,Wang:2022jwh} or the numerical solution of the JIMWLK equation \cite{Mantysaari:2018zdd}, especially in relation to the DGLAP equation \cite{Ball:1994kc}. 

In particular, the analytical solution of the BK equation with running coupling can also be used to explore the core aspects of this work. The modified BK equation with running coupling differs in form from the fixed coupling case, thus requiring the determination of a new set of parameters using the available proton structure function data. According to the discussion in Ref. \cite{Wang:2022jwh}, we find that, although the form of the solution differs slightly, both fixed and running couplings can accurately describe the existing data for the cross-sections of vector meson diffraction production. We think that the difference between these two approaches does not affect the main conclusions of this work.

\section{Summary}
\label{sec:conclu}
This work primarily utilizes the analytical solution of the BK equation \cite{Wang:2020stj} to calculate the proton structure function $F_2$ in the higher $Q^2$ region and to investigate the behavior of the structure function in the small $x$ region. We observe that as $Q^2$ increases, the slope parameter $\lambda$ (i.e., the rate of increase of $F_2$) gradually decreases and may converge at a certain value. Specifically, it approaches convergence around $\lambda \simeq 0.41 \pm 0.01$ (requiring a large $Q^2$ scale). On the other hand, we conduct a phenomenological analysis of the results and suggest that gluon overlap and phase space growth suppression can explain the behavior of the parameter $\lambda$. 

We acknowledge that the current results are highly dependent on the approximations involved in the BK equation, particularly within the LO approximation and the fixed coupling constant approximation. While our findings currently exhibit discrepancies with experimental data, phenomenological studies of high-density gluon overlap and recombination suggest the expectation of gluon saturation characteristics in a high $Q^2$ region. However, this conclusion still requires validation through future high-precision electron-ion collision experiments \cite{Accardi:2012qut,AbdulKhalek:2021gbh,Chen:2018wyz,Chen:2020ijn,Anderle:2021wcy} across a broad range of $Q^2$ and $x$.

\begin{acknowledgments}
The authors are very grateful to
Dr. Rong Wang for providing suggestions about the er-
ror analysis and fruitful discussions. This work is sup-
ported by the National Key R$\&$D Program of China
(Grant Nos. 2024YFE0109800 and 2024YFE0109802), the
National Natural Science Foundation of China (Grant
No. 12305127), and the International Partnership Pro-
gram of the Chinese Academy of Sciences (Grant
No. 016GJHZ2022054FN).
\end{acknowledgments}

\bibliography{refs}

\begin{thebibliography}{72}%
\makeatletter
\providecommand \@ifxundefined [1]{%
 \@ifx{#1\undefined}
}%
\providecommand \@ifnum [1]{%
 \ifnum #1\expandafter \@firstoftwo
 \else \expandafter \@secondoftwo
 \fi
}%
\providecommand \@ifx [1]{%
 \ifx #1\expandafter \@firstoftwo
 \else \expandafter \@secondoftwo
 \fi
}%
\providecommand \natexlab [1]{#1}%
\providecommand \enquote  [1]{``#1''}%
\providecommand \bibnamefont  [1]{#1}%
\providecommand \bibfnamefont [1]{#1}%
\providecommand \citenamefont [1]{#1}%
\providecommand \href@noop [0]{\@secondoftwo}%
\providecommand \href [0]{\begingroup \@sanitize@url \@href}%
\providecommand \@href[1]{\@@startlink{#1}\@@href}%
\providecommand \@@href[1]{\endgroup#1\@@endlink}%
\providecommand \@sanitize@url [0]{\catcode `\\12\catcode `\$12\catcode
  `\&12\catcode `\#12\catcode `\^12\catcode `\_12\catcode `\%12\relax}%
\providecommand \@@startlink[1]{}%
\providecommand \@@endlink[0]{}%
\providecommand \url  [0]{\begingroup\@sanitize@url \@url }%
\providecommand \@url [1]{\endgroup\@href {#1}{\urlprefix }}%
\providecommand \urlprefix  [0]{URL }%
\providecommand \Eprint [0]{\href }%
\providecommand \doibase [0]{https://doi.org/}%
\providecommand \selectlanguage [0]{\@gobble}%
\providecommand \bibinfo  [0]{\@secondoftwo}%
\providecommand \bibfield  [0]{\@secondoftwo}%
\providecommand \translation [1]{[#1]}%
\providecommand \BibitemOpen [0]{}%
\providecommand \bibitemStop [0]{}%
\providecommand \bibitemNoStop [0]{.\EOS\space}%
\providecommand \EOS [0]{\spacefactor3000\relax}%
\providecommand \BibitemShut  [1]{\csname bibitem#1\endcsname}%
\let\auto@bib@innerbib\@empty
\bibitem [{\citenamefont {Lipatov}(1976)}]{Lipatov:1976zz}%
  \BibitemOpen
  \bibfield  {author} {\bibinfo {author} {\bibfnamefont {L.~N.}\ \bibnamefont
  {Lipatov}},\ }\bibfield  {title} {\bibinfo {title} {{Reggeization of the
  Vector Meson and the Vacuum Singularity in Nonabelian Gauge Theories}},\
  }\href@noop {} {\bibfield  {journal} {\bibinfo  {journal} {Sov. J. Nucl.
  Phys.}\ }\textbf {\bibinfo {volume} {23}},\ \bibinfo {pages} {338} (\bibinfo
  {year} {1976})}\BibitemShut {NoStop}%
\bibitem [{\citenamefont {Kuraev}\ \emph {et~al.}(1977)\citenamefont {Kuraev},
  \citenamefont {Lipatov},\ and\ \citenamefont {Fadin}}]{Kuraev:1977fs}%
  \BibitemOpen
  \bibfield  {author} {\bibinfo {author} {\bibfnamefont {E.~A.}\ \bibnamefont
  {Kuraev}}, \bibinfo {author} {\bibfnamefont {L.~N.}\ \bibnamefont
  {Lipatov}},\ and\ \bibinfo {author} {\bibfnamefont {V.~S.}\ \bibnamefont
  {Fadin}},\ }\bibfield  {title} {\bibinfo {title} {{The Pomeranchuk
  Singularity in Nonabelian Gauge Theories}},\ }\href@noop {} {\bibfield
  {journal} {\bibinfo  {journal} {Sov. Phys. JETP}\ }\textbf {\bibinfo {volume}
  {45}},\ \bibinfo {pages} {199} (\bibinfo {year} {1977})}\BibitemShut
  {NoStop}%
\bibitem [{\citenamefont {Balitsky}\ and\ \citenamefont
  {Lipatov}(1978)}]{Balitsky:1978ic}%
  \BibitemOpen
  \bibfield  {author} {\bibinfo {author} {\bibfnamefont {I.~I.}\ \bibnamefont
  {Balitsky}}\ and\ \bibinfo {author} {\bibfnamefont {L.~N.}\ \bibnamefont
  {Lipatov}},\ }\bibfield  {title} {\bibinfo {title} {{The Pomeranchuk
  Singularity in Quantum Chromodynamics}},\ }\href@noop {} {\bibfield
  {journal} {\bibinfo  {journal} {Sov. J. Nucl. Phys.}\ }\textbf {\bibinfo
  {volume} {28}},\ \bibinfo {pages} {822} (\bibinfo {year} {1978})}\BibitemShut
  {NoStop}%
\bibitem [{\citenamefont {Ellis}\ \emph {et~al.}(2008)\citenamefont {Ellis},
  \citenamefont {Kowalski},\ and\ \citenamefont {Ross}}]{Ellis:2008yp}%
  \BibitemOpen
  \bibfield  {author} {\bibinfo {author} {\bibfnamefont {J.}~\bibnamefont
  {Ellis}}, \bibinfo {author} {\bibfnamefont {H.}~\bibnamefont {Kowalski}},\
  and\ \bibinfo {author} {\bibfnamefont {D.~A.}\ \bibnamefont {Ross}},\
  }\bibfield  {title} {\bibinfo {title} {{Evidence for the Discrete
  Asymptotically-Free BFKL Pomeron from HERA Data}},\ }\href
  {https://doi.org/10.1016/j.physletb.2008.08.007} {\bibfield  {journal}
  {\bibinfo  {journal} {Phys. Lett. B}\ }\textbf {\bibinfo {volume} {668}},\
  \bibinfo {pages} {51} (\bibinfo {year} {2008})},\ \Eprint
  {https://arxiv.org/abs/0803.0258} {arXiv:0803.0258 [hep-ph]} \BibitemShut
  {NoStop}%
\bibitem [{\citenamefont {Kowalski}\ \emph {et~al.}(2010)\citenamefont
  {Kowalski}, \citenamefont {Lipatov}, \citenamefont {Ross},\ and\
  \citenamefont {Watt}}]{Kowalski:2010ue}%
  \BibitemOpen
  \bibfield  {author} {\bibinfo {author} {\bibfnamefont {H.}~\bibnamefont
  {Kowalski}}, \bibinfo {author} {\bibfnamefont {L.~N.}\ \bibnamefont
  {Lipatov}}, \bibinfo {author} {\bibfnamefont {D.~A.}\ \bibnamefont {Ross}},\
  and\ \bibinfo {author} {\bibfnamefont {G.}~\bibnamefont {Watt}},\ }\bibfield
  {title} {\bibinfo {title} {{Using HERA Data to Determine the Infrared
  Behaviour of the BFKL Amplitude}},\ }\href
  {https://doi.org/10.1140/epjc/s10052-010-1500-6} {\bibfield  {journal}
  {\bibinfo  {journal} {Eur. Phys. J. C}\ }\textbf {\bibinfo {volume} {70}},\
  \bibinfo {pages} {983} (\bibinfo {year} {2010})},\ \Eprint
  {https://arxiv.org/abs/1005.0355} {arXiv:1005.0355 [hep-ph]} \BibitemShut
  {NoStop}%
\bibitem [{\citenamefont {Kowalski}\ \emph {et~al.}(2011)\citenamefont
  {Kowalski}, \citenamefont {Lipatov}, \citenamefont {Ross},\ and\
  \citenamefont {Watt}}]{Kowalski:2011zza}%
  \BibitemOpen
  \bibfield  {author} {\bibinfo {author} {\bibfnamefont {H.}~\bibnamefont
  {Kowalski}}, \bibinfo {author} {\bibfnamefont {L.~N.}\ \bibnamefont
  {Lipatov}}, \bibinfo {author} {\bibfnamefont {D.~A.}\ \bibnamefont {Ross}},\
  and\ \bibinfo {author} {\bibfnamefont {G.}~\bibnamefont {Watt}},\ }\bibfield
  {title} {\bibinfo {title} {{The new HERA data and the determination of the
  infrared behaviour of the BFKL amplitude}},\ }\href
  {https://doi.org/10.1016/j.nuclphysa.2010.10.002} {\bibfield  {journal}
  {\bibinfo  {journal} {Nucl. Phys. A}\ }\textbf {\bibinfo {volume} {854}},\
  \bibinfo {pages} {45} (\bibinfo {year} {2011})}\BibitemShut {NoStop}%
\bibitem [{\citenamefont {Kowalski}\ \emph {et~al.}(2013)\citenamefont
  {Kowalski}, \citenamefont {Lipatov},\ and\ \citenamefont
  {Ross}}]{Kowalski:2012ur}%
  \BibitemOpen
  \bibfield  {author} {\bibinfo {author} {\bibfnamefont {H.}~\bibnamefont
  {Kowalski}}, \bibinfo {author} {\bibfnamefont {L.~N.}\ \bibnamefont
  {Lipatov}},\ and\ \bibinfo {author} {\bibfnamefont {D.~A.}\ \bibnamefont
  {Ross}},\ }\bibfield  {title} {\bibinfo {title} {{BFKL Evolution as a
  Communicator Between Small and Large Energy Scales}},\ }\href
  {https://doi.org/10.1134/S1063779613030052} {\bibfield  {journal} {\bibinfo
  {journal} {Phys. Part. Nucl.}\ }\textbf {\bibinfo {volume} {44}},\ \bibinfo
  {pages} {547} (\bibinfo {year} {2013})},\ \Eprint
  {https://arxiv.org/abs/1205.6713} {arXiv:1205.6713 [hep-ph]} \BibitemShut
  {NoStop}%
\bibitem [{\citenamefont {Kowalski}\ \emph {et~al.}(2014)\citenamefont
  {Kowalski}, \citenamefont {Lipatov},\ and\ \citenamefont
  {Ross}}]{Kowalski:2014iqa}%
  \BibitemOpen
  \bibfield  {author} {\bibinfo {author} {\bibfnamefont {H.}~\bibnamefont
  {Kowalski}}, \bibinfo {author} {\bibfnamefont {L.}~\bibnamefont {Lipatov}},\
  and\ \bibinfo {author} {\bibfnamefont {D.}~\bibnamefont {Ross}},\ }\bibfield
  {title} {\bibinfo {title} {{The Green Function for the BFKL Pomeron and the
  Transition to DGLAP Evolution}},\ }\href
  {https://doi.org/10.1140/epjc/s10052-014-2919-y} {\bibfield  {journal}
  {\bibinfo  {journal} {Eur. Phys. J. C}\ }\textbf {\bibinfo {volume} {74}},\
  \bibinfo {pages} {2919} (\bibinfo {year} {2014})},\ \Eprint
  {https://arxiv.org/abs/1401.6298} {arXiv:1401.6298 [hep-ph]} \BibitemShut
  {NoStop}%
\bibitem [{\citenamefont {Kowalski}\ \emph {et~al.}(2016)\citenamefont
  {Kowalski}, \citenamefont {Lipatov},\ and\ \citenamefont
  {Ross}}]{Kowalski:2015paa}%
  \BibitemOpen
  \bibfield  {author} {\bibinfo {author} {\bibfnamefont {H.}~\bibnamefont
  {Kowalski}}, \bibinfo {author} {\bibfnamefont {L.~N.}\ \bibnamefont
  {Lipatov}},\ and\ \bibinfo {author} {\bibfnamefont {D.~A.}\ \bibnamefont
  {Ross}},\ }\bibfield  {title} {\bibinfo {title} {{The Behaviour of the Green
  Function for the BFKL Pomeron with Running Coupling}},\ }\href
  {https://doi.org/10.1140/epjc/s10052-015-3865-z} {\bibfield  {journal}
  {\bibinfo  {journal} {Eur. Phys. J. C}\ }\textbf {\bibinfo {volume} {76}},\
  \bibinfo {pages} {23} (\bibinfo {year} {2016})},\ \Eprint
  {https://arxiv.org/abs/1508.05744} {arXiv:1508.05744 [hep-ph]} \BibitemShut
  {NoStop}%
\bibitem [{\citenamefont {Kowalski}\ \emph {et~al.}(2017)\citenamefont
  {Kowalski}, \citenamefont {Lipatov}, \citenamefont {Ross},\ and\
  \citenamefont {Schulz}}]{Kowalski:2017umu}%
  \BibitemOpen
  \bibfield  {author} {\bibinfo {author} {\bibfnamefont {H.}~\bibnamefont
  {Kowalski}}, \bibinfo {author} {\bibfnamefont {L.~N.}\ \bibnamefont
  {Lipatov}}, \bibinfo {author} {\bibfnamefont {D.~A.}\ \bibnamefont {Ross}},\
  and\ \bibinfo {author} {\bibfnamefont {O.}~\bibnamefont {Schulz}},\
  }\bibfield  {title} {\bibinfo {title} {{Decoupling of the leading
  contribution in the discrete BFKL Analysis of High-Precision HERA Data}},\
  }\href {https://doi.org/10.1140/epjc/s10052-017-5359-7} {\bibfield  {journal}
  {\bibinfo  {journal} {Eur. Phys. J. C}\ }\textbf {\bibinfo {volume} {77}},\
  \bibinfo {pages} {777} (\bibinfo {year} {2017})},\ \Eprint
  {https://arxiv.org/abs/1707.01460} {arXiv:1707.01460 [hep-ph]} \BibitemShut
  {NoStop}%
\bibitem [{\citenamefont {Salam}(1998)}]{Salam:1998tj}%
  \BibitemOpen
  \bibfield  {author} {\bibinfo {author} {\bibfnamefont {G.~P.}\ \bibnamefont
  {Salam}},\ }\bibfield  {title} {\bibinfo {title} {{A Resummation of large
  subleading corrections at small x}},\ }\href
  {https://doi.org/10.1088/1126-6708/1998/07/019} {\bibfield  {journal}
  {\bibinfo  {journal} {JHEP}\ }\textbf {\bibinfo {volume} {07}},\ \bibinfo
  {pages} {019}},\ \Eprint {https://arxiv.org/abs/hep-ph/9806482}
  {arXiv:hep-ph/9806482} \BibitemShut {NoStop}%
\bibitem [{\citenamefont {Dokshitzer}(1977)}]{Dokshitzer:1977sg}%
  \BibitemOpen
  \bibfield  {author} {\bibinfo {author} {\bibfnamefont {Y.~L.}\ \bibnamefont
  {Dokshitzer}},\ }\bibfield  {title} {\bibinfo {title} {{Calculation of the
  Structure Functions for Deep Inelastic Scattering and e+ e- Annihilation by
  Perturbation Theory in Quantum Chromodynamics.}},\ }\href@noop {} {\bibfield
  {journal} {\bibinfo  {journal} {Sov. Phys. JETP}\ }\textbf {\bibinfo {volume}
  {46}},\ \bibinfo {pages} {641} (\bibinfo {year} {1977})}\BibitemShut
  {NoStop}%
\bibitem [{\citenamefont {Gribov}\ and\ \citenamefont
  {Lipatov}(1972)}]{Gribov:1972ri}%
  \BibitemOpen
  \bibfield  {author} {\bibinfo {author} {\bibfnamefont {V.~N.}\ \bibnamefont
  {Gribov}}\ and\ \bibinfo {author} {\bibfnamefont {L.~N.}\ \bibnamefont
  {Lipatov}},\ }\bibfield  {title} {\bibinfo {title} {{Deep inelastic e p
  scattering in perturbation theory}},\ }\href@noop {} {\bibfield  {journal}
  {\bibinfo  {journal} {Sov. J. Nucl. Phys.}\ }\textbf {\bibinfo {volume}
  {15}},\ \bibinfo {pages} {438} (\bibinfo {year} {1972})}\BibitemShut
  {NoStop}%
\bibitem [{\citenamefont {Lipatov}(1974)}]{Lipatov:1974qm}%
  \BibitemOpen
  \bibfield  {author} {\bibinfo {author} {\bibfnamefont {L.~N.}\ \bibnamefont
  {Lipatov}},\ }\bibfield  {title} {\bibinfo {title} {{The parton model and
  perturbation theory}},\ }\href@noop {} {\bibfield  {journal} {\bibinfo
  {journal} {Yad. Fiz.}\ }\textbf {\bibinfo {volume} {20}},\ \bibinfo {pages}
  {181} (\bibinfo {year} {1974})}\BibitemShut {NoStop}%
\bibitem [{\citenamefont {Altarelli}\ and\ \citenamefont
  {Parisi}(1977)}]{Altarelli:1977zs}%
  \BibitemOpen
  \bibfield  {author} {\bibinfo {author} {\bibfnamefont {G.}~\bibnamefont
  {Altarelli}}\ and\ \bibinfo {author} {\bibfnamefont {G.}~\bibnamefont
  {Parisi}},\ }\bibfield  {title} {\bibinfo {title} {{Asymptotic Freedom in
  Parton Language}},\ }\href {https://doi.org/10.1016/0550-3213(77)90384-4}
  {\bibfield  {journal} {\bibinfo  {journal} {Nucl. Phys. B}\ }\textbf
  {\bibinfo {volume} {126}},\ \bibinfo {pages} {298} (\bibinfo {year}
  {1977})}\BibitemShut {NoStop}%
\bibitem [{\citenamefont {Wilson}(1975)}]{Wilson:1974mb}%
  \BibitemOpen
  \bibfield  {author} {\bibinfo {author} {\bibfnamefont {K.~G.}\ \bibnamefont
  {Wilson}},\ }\bibfield  {title} {\bibinfo {title} {{The Renormalization
  Group: Critical Phenomena and the Kondo Problem}},\ }\href
  {https://doi.org/10.1103/RevModPhys.47.773} {\bibfield  {journal} {\bibinfo
  {journal} {Rev. Mod. Phys.}\ }\textbf {\bibinfo {volume} {47}},\ \bibinfo
  {pages} {773} (\bibinfo {year} {1975})}\BibitemShut {NoStop}%
\bibitem [{\citenamefont {Mueller}(2001)}]{Mueller:2001fv}%
  \BibitemOpen
  \bibfield  {author} {\bibinfo {author} {\bibfnamefont {A.~H.}\ \bibnamefont
  {Mueller}},\ }\bibfield  {title} {\bibinfo {title} {{Parton saturation: An
  Overview}},\ }in\ \href@noop {} {\emph {\bibinfo {booktitle} {{Cargese Summer
  School on QCD Perspectives on Hot and Dense Matter}}}}\ (\bibinfo {year}
  {2001})\ pp.\ \bibinfo {pages} {45--72},\ \Eprint
  {https://arxiv.org/abs/hep-ph/0111244} {arXiv:hep-ph/0111244} \BibitemShut
  {NoStop}%
\bibitem [{\citenamefont {Stasto}\ \emph {et~al.}(2001)\citenamefont {Stasto},
  \citenamefont {Golec-Biernat},\ and\ \citenamefont
  {Kwiecinski}}]{Stasto:2000er}%
  \BibitemOpen
  \bibfield  {author} {\bibinfo {author} {\bibfnamefont {A.~M.}\ \bibnamefont
  {Stasto}}, \bibinfo {author} {\bibfnamefont {K.~J.}\ \bibnamefont
  {Golec-Biernat}},\ and\ \bibinfo {author} {\bibfnamefont {J.}~\bibnamefont
  {Kwiecinski}},\ }\bibfield  {title} {\bibinfo {title} {{Geometric scaling for
  the total gamma* p cross-section in the low x region}},\ }\href
  {https://doi.org/10.1103/PhysRevLett.86.596} {\bibfield  {journal} {\bibinfo
  {journal} {Phys. Rev. Lett.}\ }\textbf {\bibinfo {volume} {86}},\ \bibinfo
  {pages} {596} (\bibinfo {year} {2001})},\ \Eprint
  {https://arxiv.org/abs/hep-ph/0007192} {arXiv:hep-ph/0007192} \BibitemShut
  {NoStop}%
\bibitem [{\citenamefont {Munier}\ and\ \citenamefont
  {Peschanski}(2003)}]{Munier:2003vc}%
  \BibitemOpen
  \bibfield  {author} {\bibinfo {author} {\bibfnamefont {S.}~\bibnamefont
  {Munier}}\ and\ \bibinfo {author} {\bibfnamefont {R.~B.}\ \bibnamefont
  {Peschanski}},\ }\bibfield  {title} {\bibinfo {title} {{Geometric scaling as
  traveling waves}},\ }\href {https://doi.org/10.1103/PhysRevLett.91.232001}
  {\bibfield  {journal} {\bibinfo  {journal} {Phys. Rev. Lett.}\ }\textbf
  {\bibinfo {volume} {91}},\ \bibinfo {pages} {232001} (\bibinfo {year}
  {2003})},\ \Eprint {https://arxiv.org/abs/hep-ph/0309177}
  {arXiv:hep-ph/0309177} \BibitemShut {NoStop}%
\bibitem [{\citenamefont {Balitsky}(1997)}]{Balitsky:1997mk}%
  \BibitemOpen
  \bibfield  {author} {\bibinfo {author} {\bibfnamefont {I.}~\bibnamefont
  {Balitsky}},\ }\bibfield  {title} {\bibinfo {title} {{Operator expansion for
  diffractive high-energy scattering}},\ }\href
  {https://doi.org/10.1063/1.53693} {\bibfield  {journal} {\bibinfo  {journal}
  {AIP Conf. Proc.}\ }\textbf {\bibinfo {volume} {407}},\ \bibinfo {pages}
  {953} (\bibinfo {year} {1997})},\ \Eprint
  {https://arxiv.org/abs/hep-ph/9706411} {arXiv:hep-ph/9706411} \BibitemShut
  {NoStop}%
\bibitem [{\citenamefont {Kovchegov}(1999)}]{Kovchegov:1999yj}%
  \BibitemOpen
  \bibfield  {author} {\bibinfo {author} {\bibfnamefont {Y.~V.}\ \bibnamefont
  {Kovchegov}},\ }\bibfield  {title} {\bibinfo {title} {{Small x F(2) structure
  function of a nucleus including multiple pomeron exchanges}},\ }\href
  {https://doi.org/10.1103/PhysRevD.60.034008} {\bibfield  {journal} {\bibinfo
  {journal} {Phys. Rev. D}\ }\textbf {\bibinfo {volume} {60}},\ \bibinfo
  {pages} {034008} (\bibinfo {year} {1999})},\ \Eprint
  {https://arxiv.org/abs/hep-ph/9901281} {arXiv:hep-ph/9901281} \BibitemShut
  {NoStop}%
\bibitem [{\citenamefont {Kovchegov}(2000)}]{Kovchegov:1999ua}%
  \BibitemOpen
  \bibfield  {author} {\bibinfo {author} {\bibfnamefont {Y.~V.}\ \bibnamefont
  {Kovchegov}},\ }\bibfield  {title} {\bibinfo {title} {{Unitarization of the
  BFKL pomeron on a nucleus}},\ }\href
  {https://doi.org/10.1103/PhysRevD.61.074018} {\bibfield  {journal} {\bibinfo
  {journal} {Phys. Rev. D}\ }\textbf {\bibinfo {volume} {61}},\ \bibinfo
  {pages} {074018} (\bibinfo {year} {2000})},\ \Eprint
  {https://arxiv.org/abs/hep-ph/9905214} {arXiv:hep-ph/9905214} \BibitemShut
  {NoStop}%
\bibitem [{\citenamefont {Balitsky}(2001)}]{Balitsky:2001re}%
  \BibitemOpen
  \bibfield  {author} {\bibinfo {author} {\bibfnamefont {I.}~\bibnamefont
  {Balitsky}},\ }\bibfield  {title} {\bibinfo {title} {{Effective field theory
  for the small x evolution}},\ }\href
  {https://doi.org/10.1016/S0370-2693(01)01041-3} {\bibfield  {journal}
  {\bibinfo  {journal} {Phys. Lett. B}\ }\textbf {\bibinfo {volume} {518}},\
  \bibinfo {pages} {235} (\bibinfo {year} {2001})},\ \Eprint
  {https://arxiv.org/abs/hep-ph/0105334} {arXiv:hep-ph/0105334} \BibitemShut
  {NoStop}%
\bibitem [{\citenamefont {Balitsky}(1996)}]{Balitsky:1995ub}%
  \BibitemOpen
  \bibfield  {author} {\bibinfo {author} {\bibfnamefont {I.}~\bibnamefont
  {Balitsky}},\ }\bibfield  {title} {\bibinfo {title} {{Operator expansion for
  high-energy scattering}},\ }\href
  {https://doi.org/10.1016/0550-3213(95)00638-9} {\bibfield  {journal}
  {\bibinfo  {journal} {Nucl. Phys. B}\ }\textbf {\bibinfo {volume} {463}},\
  \bibinfo {pages} {99} (\bibinfo {year} {1996})},\ \Eprint
  {https://arxiv.org/abs/hep-ph/9509348} {arXiv:hep-ph/9509348} \BibitemShut
  {NoStop}%
\bibitem [{\citenamefont {Jalilian-Marian}\ \emph {et~al.}(1997)\citenamefont
  {Jalilian-Marian}, \citenamefont {Kovner}, \citenamefont {Leonidov},\ and\
  \citenamefont {Weigert}}]{Jalilian-Marian:1997qno}%
  \BibitemOpen
  \bibfield  {author} {\bibinfo {author} {\bibfnamefont {J.}~\bibnamefont
  {Jalilian-Marian}}, \bibinfo {author} {\bibfnamefont {A.}~\bibnamefont
  {Kovner}}, \bibinfo {author} {\bibfnamefont {A.}~\bibnamefont {Leonidov}},\
  and\ \bibinfo {author} {\bibfnamefont {H.}~\bibnamefont {Weigert}},\
  }\bibfield  {title} {\bibinfo {title} {{The BFKL equation from the Wilson
  renormalization group}},\ }\href
  {https://doi.org/10.1016/S0550-3213(97)00440-9} {\bibfield  {journal}
  {\bibinfo  {journal} {Nucl. Phys. B}\ }\textbf {\bibinfo {volume} {504}},\
  \bibinfo {pages} {415} (\bibinfo {year} {1997})},\ \Eprint
  {https://arxiv.org/abs/hep-ph/9701284} {arXiv:hep-ph/9701284} \BibitemShut
  {NoStop}%
\bibitem [{\citenamefont {Iancu}\ \emph {et~al.}(2001)\citenamefont {Iancu},
  \citenamefont {Leonidov},\ and\ \citenamefont {McLerran}}]{Iancu:2000hn}%
  \BibitemOpen
  \bibfield  {author} {\bibinfo {author} {\bibfnamefont {E.}~\bibnamefont
  {Iancu}}, \bibinfo {author} {\bibfnamefont {A.}~\bibnamefont {Leonidov}},\
  and\ \bibinfo {author} {\bibfnamefont {L.~D.}\ \bibnamefont {McLerran}},\
  }\bibfield  {title} {\bibinfo {title} {{Nonlinear gluon evolution in the
  color glass condensate. 1.}},\ }\href
  {https://doi.org/10.1016/S0375-9474(01)00642-X} {\bibfield  {journal}
  {\bibinfo  {journal} {Nucl. Phys. A}\ }\textbf {\bibinfo {volume} {692}},\
  \bibinfo {pages} {583} (\bibinfo {year} {2001})},\ \Eprint
  {https://arxiv.org/abs/hep-ph/0011241} {arXiv:hep-ph/0011241} \BibitemShut
  {NoStop}%
\bibitem [{\citenamefont {Weigert}(2002)}]{Weigert:2000gi}%
  \BibitemOpen
  \bibfield  {author} {\bibinfo {author} {\bibfnamefont {H.}~\bibnamefont
  {Weigert}},\ }\bibfield  {title} {\bibinfo {title} {{Unitarity at small
  Bjorken x}},\ }\href {https://doi.org/10.1016/S0375-9474(01)01668-2}
  {\bibfield  {journal} {\bibinfo  {journal} {Nucl. Phys. A}\ }\textbf
  {\bibinfo {volume} {703}},\ \bibinfo {pages} {823} (\bibinfo {year}
  {2002})},\ \Eprint {https://arxiv.org/abs/hep-ph/0004044}
  {arXiv:hep-ph/0004044} \BibitemShut {NoStop}%
\bibitem [{\citenamefont {Wang}\ \emph {et~al.}(2021)\citenamefont {Wang},
  \citenamefont {Yang}, \citenamefont {Kou}, \citenamefont {Wang},\ and\
  \citenamefont {Chen}}]{Wang:2020stj}%
  \BibitemOpen
  \bibfield  {author} {\bibinfo {author} {\bibfnamefont {X.}~\bibnamefont
  {Wang}}, \bibinfo {author} {\bibfnamefont {Y.}~\bibnamefont {Yang}}, \bibinfo
  {author} {\bibfnamefont {W.}~\bibnamefont {Kou}}, \bibinfo {author}
  {\bibfnamefont {R.}~\bibnamefont {Wang}},\ and\ \bibinfo {author}
  {\bibfnamefont {X.}~\bibnamefont {Chen}},\ }\bibfield  {title} {\bibinfo
  {title} {{Analytical solution of Balitsky-Kovchegov equation with homogeneous
  balance method}},\ }\href {https://doi.org/10.1103/PhysRevD.103.056008}
  {\bibfield  {journal} {\bibinfo  {journal} {Phys. Rev. D}\ }\textbf {\bibinfo
  {volume} {103}},\ \bibinfo {pages} {056008} (\bibinfo {year} {2021})},\
  \Eprint {https://arxiv.org/abs/2009.13325} {arXiv:2009.13325 [hep-ph]}
  \BibitemShut {NoStop}%
\bibitem [{\citenamefont {Bartels}\ and\ \citenamefont
  {Kowalski}(2001)}]{Bartels:2000ze}%
  \BibitemOpen
  \bibfield  {author} {\bibinfo {author} {\bibfnamefont {J.}~\bibnamefont
  {Bartels}}\ and\ \bibinfo {author} {\bibfnamefont {H.}~\bibnamefont
  {Kowalski}},\ }\bibfield  {title} {\bibinfo {title} {{Diffraction at HERA and
  the confinement problem}},\ }\href {https://doi.org/10.1007/s100520100613}
  {\bibfield  {journal} {\bibinfo  {journal} {Eur. Phys. J. C}\ }\textbf
  {\bibinfo {volume} {19}},\ \bibinfo {pages} {693} (\bibinfo {year} {2001})},\
  \Eprint {https://arxiv.org/abs/hep-ph/0010345} {arXiv:hep-ph/0010345}
  \BibitemShut {NoStop}%
\bibitem [{\citenamefont {Cvetic}\ \emph {et~al.}(2009)\citenamefont {Cvetic},
  \citenamefont {Illarionov}, \citenamefont {Kniehl},\ and\ \citenamefont
  {Kotikov}}]{Cvetic:2009kw}%
  \BibitemOpen
  \bibfield  {author} {\bibinfo {author} {\bibfnamefont {G.}~\bibnamefont
  {Cvetic}}, \bibinfo {author} {\bibfnamefont {A.~Y.}\ \bibnamefont
  {Illarionov}}, \bibinfo {author} {\bibfnamefont {B.~A.}\ \bibnamefont
  {Kniehl}},\ and\ \bibinfo {author} {\bibfnamefont {A.~V.}\ \bibnamefont
  {Kotikov}},\ }\bibfield  {title} {\bibinfo {title} {{Small-x behavior of the
  structure function F(2) and its slope partial ln F(2) / partial ln(1/x) for
  'frozen' and analytic strong-coupling constants}},\ }\href
  {https://doi.org/10.1016/j.physletb.2009.07.057} {\bibfield  {journal}
  {\bibinfo  {journal} {Phys. Lett. B}\ }\textbf {\bibinfo {volume} {679}},\
  \bibinfo {pages} {350} (\bibinfo {year} {2009})},\ \Eprint
  {https://arxiv.org/abs/0906.1925} {arXiv:0906.1925 [hep-ph]} \BibitemShut
  {NoStop}%
\bibitem [{\citenamefont {Cooper-Sarkar}\ \emph {et~al.}(1998)\citenamefont
  {Cooper-Sarkar}, \citenamefont {Devenish},\ and\ \citenamefont
  {De~Roeck}}]{Cooper-Sarkar:1997pqx}%
  \BibitemOpen
  \bibfield  {author} {\bibinfo {author} {\bibfnamefont {A.~M.}\ \bibnamefont
  {Cooper-Sarkar}}, \bibinfo {author} {\bibfnamefont {R.~C.~E.}\ \bibnamefont
  {Devenish}},\ and\ \bibinfo {author} {\bibfnamefont {A.}~\bibnamefont
  {De~Roeck}},\ }\bibfield  {title} {\bibinfo {title} {{Structure functions of
  the nucleon and their interpretation}},\ }\href
  {https://doi.org/10.1142/S0217751X98001670} {\bibfield  {journal} {\bibinfo
  {journal} {Int. J. Mod. Phys. A}\ }\textbf {\bibinfo {volume} {13}},\
  \bibinfo {pages} {3385} (\bibinfo {year} {1998})},\ \Eprint
  {https://arxiv.org/abs/hep-ph/9712301} {arXiv:hep-ph/9712301} \BibitemShut
  {NoStop}%
\bibitem [{\citenamefont {Kotikov}\ and\ \citenamefont
  {Parente}(1999)}]{Kotikov:1998qt}%
  \BibitemOpen
  \bibfield  {author} {\bibinfo {author} {\bibfnamefont {A.~V.}\ \bibnamefont
  {Kotikov}}\ and\ \bibinfo {author} {\bibfnamefont {G.}~\bibnamefont
  {Parente}},\ }\bibfield  {title} {\bibinfo {title} {{Small x behavior of
  parton distributions with soft initial conditions}},\ }\href
  {https://doi.org/10.1016/S0550-3213(99)00107-8} {\bibfield  {journal}
  {\bibinfo  {journal} {Nucl. Phys. B}\ }\textbf {\bibinfo {volume} {549}},\
  \bibinfo {pages} {242} (\bibinfo {year} {1999})},\ \Eprint
  {https://arxiv.org/abs/hep-ph/9807249} {arXiv:hep-ph/9807249} \BibitemShut
  {NoStop}%
\bibitem [{\citenamefont {Kotikov}(2007)}]{Kotikov:2007ua}%
  \BibitemOpen
  \bibfield  {author} {\bibinfo {author} {\bibfnamefont {A.~V.}\ \bibnamefont
  {Kotikov}},\ }\bibfield  {title} {\bibinfo {title} {{Deep inelastic
  scattering: Q**2 dependence of structure functions}},\ }\href
  {https://doi.org/10.1134/S1063779607010017} {\bibfield  {journal} {\bibinfo
  {journal} {Phys. Part. Nucl.}\ }\textbf {\bibinfo {volume} {38}},\ \bibinfo
  {pages} {1} (\bibinfo {year} {2007})},\ \bibinfo {note} {[Erratum:
  Phys.Part.Nucl. 38, 828--829 (2007)]}\BibitemShut {NoStop}%
\bibitem [{\citenamefont {Ball}\ and\ \citenamefont
  {Forte}(1994)}]{Ball:1994kc}%
  \BibitemOpen
  \bibfield  {author} {\bibinfo {author} {\bibfnamefont {R.~D.}\ \bibnamefont
  {Ball}}\ and\ \bibinfo {author} {\bibfnamefont {S.}~\bibnamefont {Forte}},\
  }\bibfield  {title} {\bibinfo {title} {{A Direct test of perturbative QCD at
  small x}},\ }\href {https://doi.org/10.1016/0370-2693(94)00956-2} {\bibfield
  {journal} {\bibinfo  {journal} {Phys. Lett. B}\ }\textbf {\bibinfo {volume}
  {336}},\ \bibinfo {pages} {77} (\bibinfo {year} {1994})},\ \Eprint
  {https://arxiv.org/abs/hep-ph/9406385} {arXiv:hep-ph/9406385} \BibitemShut
  {NoStop}%
\bibitem [{\citenamefont {Illarionov}\ \emph {et~al.}(2008)\citenamefont
  {Illarionov}, \citenamefont {Kotikov},\ and\ \citenamefont
  {Parente~Bermudez}}]{Illarionov:2004nw}%
  \BibitemOpen
  \bibfield  {author} {\bibinfo {author} {\bibfnamefont {A.~Y.}\ \bibnamefont
  {Illarionov}}, \bibinfo {author} {\bibfnamefont {A.~V.}\ \bibnamefont
  {Kotikov}},\ and\ \bibinfo {author} {\bibfnamefont {G.}~\bibnamefont
  {Parente~Bermudez}},\ }\bibfield  {title} {\bibinfo {title} {{Small x
  behavior of parton distributions. A Study of higher twist effects}},\ }\href
  {https://doi.org/10.1134/S1063779608030015} {\bibfield  {journal} {\bibinfo
  {journal} {Phys. Part. Nucl.}\ }\textbf {\bibinfo {volume} {39}},\ \bibinfo
  {pages} {307} (\bibinfo {year} {2008})},\ \Eprint
  {https://arxiv.org/abs/hep-ph/0402173} {arXiv:hep-ph/0402173} \BibitemShut
  {NoStop}%
\bibitem [{\citenamefont {Mankiewicz}\ \emph {et~al.}(1997)\citenamefont
  {Mankiewicz}, \citenamefont {Saalfeld},\ and\ \citenamefont
  {Weigl}}]{Mankiewicz:1996sd}%
  \BibitemOpen
  \bibfield  {author} {\bibinfo {author} {\bibfnamefont {L.}~\bibnamefont
  {Mankiewicz}}, \bibinfo {author} {\bibfnamefont {A.}~\bibnamefont
  {Saalfeld}},\ and\ \bibinfo {author} {\bibfnamefont {T.}~\bibnamefont
  {Weigl}},\ }\bibfield  {title} {\bibinfo {title} {{On the analytical
  approximation to the GLAP evolution at small x and moderate Q**2}},\ }\href
  {https://doi.org/10.1016/S0370-2693(96)01615-2} {\bibfield  {journal}
  {\bibinfo  {journal} {Phys. Lett. B}\ }\textbf {\bibinfo {volume} {393}},\
  \bibinfo {pages} {175} (\bibinfo {year} {1997})},\ \Eprint
  {https://arxiv.org/abs/hep-ph/9612297} {arXiv:hep-ph/9612297} \BibitemShut
  {NoStop}%
\bibitem [{\citenamefont {Abramowicz}\ \emph {et~al.}(2015)\citenamefont
  {Abramowicz} \emph {et~al.}}]{H1:2015ubc}%
  \BibitemOpen
  \bibfield  {author} {\bibinfo {author} {\bibfnamefont {H.}~\bibnamefont
  {Abramowicz}} \emph {et~al.} (\bibinfo {collaboration} {H1, ZEUS}),\
  }\bibfield  {title} {\bibinfo {title} {{Combination of measurements of
  inclusive deep inelastic ${e^{\pm }p}$ scattering cross sections and QCD
  analysis of HERA data}},\ }\href
  {https://doi.org/10.1140/epjc/s10052-015-3710-4} {\bibfield  {journal}
  {\bibinfo  {journal} {Eur. Phys. J. C}\ }\textbf {\bibinfo {volume} {75}},\
  \bibinfo {pages} {580} (\bibinfo {year} {2015})},\ \Eprint
  {https://arxiv.org/abs/1506.06042} {arXiv:1506.06042 [hep-ex]} \BibitemShut
  {NoStop}%
\bibitem [{\citenamefont {Luszczak}\ and\ \citenamefont
  {Kowalski}(2020)}]{Luszczak:2019dsp}%
  \BibitemOpen
  \bibfield  {author} {\bibinfo {author} {\bibfnamefont {A.}~\bibnamefont
  {Luszczak}}\ and\ \bibinfo {author} {\bibfnamefont {H.}~\bibnamefont
  {Kowalski}},\ }\bibfield  {title} {\bibinfo {title} {{Investigation of High
  Energy Behaviour of HERA Data}},\ }\href
  {https://doi.org/10.1016/j.physletb.2020.135199} {\bibfield  {journal}
  {\bibinfo  {journal} {Phys. Lett. B}\ }\textbf {\bibinfo {volume} {802}},\
  \bibinfo {pages} {135199} (\bibinfo {year} {2020})},\ \Eprint
  {https://arxiv.org/abs/1903.09719} {arXiv:1903.09719 [hep-ph]} \BibitemShut
  {NoStop}%
\bibitem [{\citenamefont {Kaidalov}\ \emph {et~al.}(2001)\citenamefont
  {Kaidalov}, \citenamefont {Merino},\ and\ \citenamefont
  {Pertermann}}]{Kaidalov:2000wg}%
  \BibitemOpen
  \bibfield  {author} {\bibinfo {author} {\bibfnamefont {A.~B.}\ \bibnamefont
  {Kaidalov}}, \bibinfo {author} {\bibfnamefont {C.}~\bibnamefont {Merino}},\
  and\ \bibinfo {author} {\bibfnamefont {D.}~\bibnamefont {Pertermann}},\
  }\bibfield  {title} {\bibinfo {title} {{On the behavior of F(2) and its
  logarithmic slopes}},\ }\href {https://doi.org/10.1007/s100520100648}
  {\bibfield  {journal} {\bibinfo  {journal} {Eur. Phys. J. C}\ }\textbf
  {\bibinfo {volume} {20}},\ \bibinfo {pages} {301} (\bibinfo {year} {2001})},\
  \Eprint {https://arxiv.org/abs/hep-ph/0004237} {arXiv:hep-ph/0004237}
  \BibitemShut {NoStop}%
\bibitem [{\citenamefont {Donnachie}\ and\ \citenamefont
  {Landshoff}(2003)}]{Donnachie:2003cs}%
  \BibitemOpen
  \bibfield  {author} {\bibinfo {author} {\bibfnamefont {A.}~\bibnamefont
  {Donnachie}}\ and\ \bibinfo {author} {\bibfnamefont {P.~V.}\ \bibnamefont
  {Landshoff}},\ }\bibfield  {title} {\bibinfo {title} {{Evolution at small
  x}},\ }\href@noop {} {\bibfield  {journal} {\bibinfo  {journal} {Acta Phys.
  Polon. B}\ }\textbf {\bibinfo {volume} {34}},\ \bibinfo {pages} {2989}
  (\bibinfo {year} {2003})},\ \Eprint {https://arxiv.org/abs/hep-ph/0305171}
  {arXiv:hep-ph/0305171} \BibitemShut {NoStop}%
\bibitem [{\citenamefont {Mueller}\ and\ \citenamefont
  {Patel}(1994)}]{Mueller:1994jq}%
  \BibitemOpen
  \bibfield  {author} {\bibinfo {author} {\bibfnamefont {A.~H.}\ \bibnamefont
  {Mueller}}\ and\ \bibinfo {author} {\bibfnamefont {B.}~\bibnamefont
  {Patel}},\ }\bibfield  {title} {\bibinfo {title} {{Single and double BFKL
  pomeron exchange and a dipole picture of high-energy hard processes}},\
  }\href {https://doi.org/10.1016/0550-3213(94)90284-4} {\bibfield  {journal}
  {\bibinfo  {journal} {Nucl. Phys. B}\ }\textbf {\bibinfo {volume} {425}},\
  \bibinfo {pages} {471} (\bibinfo {year} {1994})},\ \Eprint
  {https://arxiv.org/abs/hep-ph/9403256} {arXiv:hep-ph/9403256} \BibitemShut
  {NoStop}%
\bibitem [{\citenamefont {Kowalski}\ and\ \citenamefont
  {Teaney}(2003)}]{Kowalski:2003hm}%
  \BibitemOpen
  \bibfield  {author} {\bibinfo {author} {\bibfnamefont {H.}~\bibnamefont
  {Kowalski}}\ and\ \bibinfo {author} {\bibfnamefont {D.}~\bibnamefont
  {Teaney}},\ }\bibfield  {title} {\bibinfo {title} {{An Impact parameter
  dipole saturation model}},\ }\href
  {https://doi.org/10.1103/PhysRevD.68.114005} {\bibfield  {journal} {\bibinfo
  {journal} {Phys. Rev. D}\ }\textbf {\bibinfo {volume} {68}},\ \bibinfo
  {pages} {114005} (\bibinfo {year} {2003})},\ \Eprint
  {https://arxiv.org/abs/hep-ph/0304189} {arXiv:hep-ph/0304189} \BibitemShut
  {NoStop}%
\bibitem [{\citenamefont {Kowalski}\ \emph {et~al.}(2006)\citenamefont
  {Kowalski}, \citenamefont {Motyka},\ and\ \citenamefont
  {Watt}}]{Kowalski:2006hc}%
  \BibitemOpen
  \bibfield  {author} {\bibinfo {author} {\bibfnamefont {H.}~\bibnamefont
  {Kowalski}}, \bibinfo {author} {\bibfnamefont {L.}~\bibnamefont {Motyka}},\
  and\ \bibinfo {author} {\bibfnamefont {G.}~\bibnamefont {Watt}},\ }\bibfield
  {title} {\bibinfo {title} {{Exclusive diffractive processes at HERA within
  the dipole picture}},\ }\href {https://doi.org/10.1103/PhysRevD.74.074016}
  {\bibfield  {journal} {\bibinfo  {journal} {Phys. Rev. D}\ }\textbf {\bibinfo
  {volume} {74}},\ \bibinfo {pages} {074016} (\bibinfo {year} {2006})},\
  \Eprint {https://arxiv.org/abs/hep-ph/0606272} {arXiv:hep-ph/0606272}
  \BibitemShut {NoStop}%
\bibitem [{\citenamefont {de~Santana~Amaral}\ \emph {et~al.}(2007)\citenamefont
  {de~Santana~Amaral}, \citenamefont {Gay~Ducati}, \citenamefont {Betemps},\
  and\ \citenamefont {Soyez}}]{deSantanaAmaral:2006fe}%
  \BibitemOpen
  \bibfield  {author} {\bibinfo {author} {\bibfnamefont {J.~T.}\ \bibnamefont
  {de~Santana~Amaral}}, \bibinfo {author} {\bibfnamefont {M.~B.}\ \bibnamefont
  {Gay~Ducati}}, \bibinfo {author} {\bibfnamefont {M.~A.}\ \bibnamefont
  {Betemps}},\ and\ \bibinfo {author} {\bibfnamefont {G.}~\bibnamefont
  {Soyez}},\ }\bibfield  {title} {\bibinfo {title} {{gamma* p cross-section
  from the dipole model in momentum space}},\ }\href
  {https://doi.org/10.1103/PhysRevD.76.094018} {\bibfield  {journal} {\bibinfo
  {journal} {Phys. Rev. D}\ }\textbf {\bibinfo {volume} {76}},\ \bibinfo
  {pages} {094018} (\bibinfo {year} {2007})},\ \Eprint
  {https://arxiv.org/abs/hep-ph/0612091} {arXiv:hep-ph/0612091} \BibitemShut
  {NoStop}%
\bibitem [{\citenamefont {Barone}\ and\ \citenamefont
  {Predazzi}(2002)}]{Barone:2002cv}%
  \BibitemOpen
  \bibfield  {author} {\bibinfo {author} {\bibfnamefont {V.}~\bibnamefont
  {Barone}}\ and\ \bibinfo {author} {\bibfnamefont {E.}~\bibnamefont
  {Predazzi}},\ }\href@noop {} {\emph {\bibinfo {title} {{High-Energy Particle
  Diffraction}}}},\ \bibinfo {series} {Texts and Monographs in Physics}, Vol.\
  \bibinfo {volume} {v.565}\ (\bibinfo  {publisher} {Springer-Verlag},\
  \bibinfo {address} {Berlin Heidelberg},\ \bibinfo {year} {2002})\BibitemShut
  {NoStop}%
\bibitem [{\citenamefont {Xie}\ and\ \citenamefont {Chen}(2018)}]{Xie2015}%
  \BibitemOpen
  \bibfield  {author} {\bibinfo {author} {\bibfnamefont {Y.-P.}\ \bibnamefont
  {Xie}}\ and\ \bibinfo {author} {\bibfnamefont {X.}~\bibnamefont {Chen}},\
  }\bibfield  {title} {\bibinfo {title} {Exclusive j/$\psi$ photoproduction in
  a diffractive process using the ads/qcd holographic wave function in blfq},\
  }\href {https://doi.org/10.1142/S0217751X18500343} {\bibfield  {journal}
  {\bibinfo  {journal} {International Journal of Modern Physics A}\ }\textbf
  {\bibinfo {volume} {33}},\ \bibinfo {pages} {1850034} (\bibinfo {year}
  {2018})},\ \Eprint
  {https://arxiv.org/abs/https://doi.org/10.1142/S0217751X18500343}
  {https://doi.org/10.1142/S0217751X18500343} \BibitemShut {NoStop}%
\bibitem [{\citenamefont
  {FISHER}(1937)}]{https://doi.org/10.1111/j.1469-1809.1937.tb02153.x}%
  \BibitemOpen
  \bibfield  {author} {\bibinfo {author} {\bibfnamefont {R.~A.}\ \bibnamefont
  {FISHER}},\ }\bibfield  {title} {\bibinfo {title} {The wave of advance of
  advantageous genes},\ }\href
  {https://doi.org/https://doi.org/10.1111/j.1469-1809.1937.tb02153.x}
  {\bibfield  {journal} {\bibinfo  {journal} {Annals of Eugenics}\ }\textbf
  {\bibinfo {volume} {7}},\ \bibinfo {pages} {355} (\bibinfo {year}
  {1937})}\BibitemShut {NoStop}%
\bibitem [{\citenamefont {Wang}(1995)}]{WANG1995169}%
  \BibitemOpen
  \bibfield  {author} {\bibinfo {author} {\bibfnamefont {M.}~\bibnamefont
  {Wang}},\ }\bibfield  {title} {\bibinfo {title} {Solitary wave solutions for
  variant boussinesq equations},\ }\href
  {https://doi.org/https://doi.org/10.1016/0375-9601(95)00092-H} {\bibfield
  {journal} {\bibinfo  {journal} {Physics Letters A}\ }\textbf {\bibinfo
  {volume} {199}},\ \bibinfo {pages} {169} (\bibinfo {year}
  {1995})}\BibitemShut {NoStop}%
\bibitem [{\citenamefont {Wang}(1996)}]{WANG1996279}%
  \BibitemOpen
  \bibfield  {author} {\bibinfo {author} {\bibfnamefont {M.}~\bibnamefont
  {Wang}},\ }\bibfield  {title} {\bibinfo {title} {Exact solutions for a
  compound kdv-burgers equation},\ }\href
  {https://doi.org/https://doi.org/10.1016/0375-9601(96)00103-X} {\bibfield
  {journal} {\bibinfo  {journal} {Physics Letters A}\ }\textbf {\bibinfo
  {volume} {213}},\ \bibinfo {pages} {279} (\bibinfo {year}
  {1996})}\BibitemShut {NoStop}%
\bibitem [{\citenamefont {Zhou}\ \emph {et~al.}(2003)\citenamefont {Zhou},
  \citenamefont {Wang},\ and\ \citenamefont {Wang}}]{ZHOU200331}%
  \BibitemOpen
  \bibfield  {author} {\bibinfo {author} {\bibfnamefont {Y.}~\bibnamefont
  {Zhou}}, \bibinfo {author} {\bibfnamefont {M.}~\bibnamefont {Wang}},\ and\
  \bibinfo {author} {\bibfnamefont {Y.}~\bibnamefont {Wang}},\ }\bibfield
  {title} {\bibinfo {title} {Periodic wave solutions to a coupled kdv equations
  with variable coefficients},\ }\href
  {https://doi.org/https://doi.org/10.1016/S0375-9601(02)01775-9} {\bibfield
  {journal} {\bibinfo  {journal} {Physics Letters A}\ }\textbf {\bibinfo
  {volume} {308}},\ \bibinfo {pages} {31} (\bibinfo {year} {2003})}\BibitemShut
  {NoStop}%
\bibitem [{\citenamefont {Zhou}\ \emph {et~al.}(2004)\citenamefont {Zhou},
  \citenamefont {Wang},\ and\ \citenamefont {Miao}}]{ZHOU200477}%
  \BibitemOpen
  \bibfield  {author} {\bibinfo {author} {\bibfnamefont {Y.}~\bibnamefont
  {Zhou}}, \bibinfo {author} {\bibfnamefont {M.}~\bibnamefont {Wang}},\ and\
  \bibinfo {author} {\bibfnamefont {T.}~\bibnamefont {Miao}},\ }\bibfield
  {title} {\bibinfo {title} {The periodic wave solutions and solitary wave
  solutions for a class of nonlinear partial differential equations},\ }\href
  {https://doi.org/https://doi.org/10.1016/j.physleta.2004.01.056} {\bibfield
  {journal} {\bibinfo  {journal} {Physics Letters A}\ }\textbf {\bibinfo
  {volume} {323}},\ \bibinfo {pages} {77} (\bibinfo {year} {2004})}\BibitemShut
  {NoStop}%
\bibitem [{\citenamefont {Wang}\ and\ \citenamefont
  {Li}(2014)}]{wang2014simplified}%
  \BibitemOpen
  \bibfield  {author} {\bibinfo {author} {\bibfnamefont {M.}~\bibnamefont
  {Wang}}\ and\ \bibinfo {author} {\bibfnamefont {X.}~\bibnamefont {Li}},\
  }\bibfield  {title} {\bibinfo {title} {Simplified homogeneous balance method
  and its applications to the whitham-broer-kaup model equations},\ }\href@noop
  {} {\bibfield  {journal} {\bibinfo  {journal} {Journal of applied mathematics
  and physics}\ }\textbf {\bibinfo {volume} {2014}} (\bibinfo {year}
  {2014})}\BibitemShut {NoStop}%
\bibitem [{\citenamefont {Marquet}\ and\ \citenamefont
  {Soyez}(2005)}]{Marquet:2005zf}%
  \BibitemOpen
  \bibfield  {author} {\bibinfo {author} {\bibfnamefont {C.}~\bibnamefont
  {Marquet}}\ and\ \bibinfo {author} {\bibfnamefont {G.}~\bibnamefont
  {Soyez}},\ }\bibfield  {title} {\bibinfo {title} {{The Balitsky-Kovchegov
  equation in full momentum space}},\ }\href
  {https://doi.org/10.1016/j.nuclphysa.2005.05.198} {\bibfield  {journal}
  {\bibinfo  {journal} {Nucl. Phys. A}\ }\textbf {\bibinfo {volume} {760}},\
  \bibinfo {pages} {208} (\bibinfo {year} {2005})},\ \Eprint
  {https://arxiv.org/abs/hep-ph/0504080} {arXiv:hep-ph/0504080} \BibitemShut
  {NoStop}%
\bibitem [{\citenamefont {Andreev}\ \emph {et~al.}(2014)\citenamefont {Andreev}
  \emph {et~al.}}]{H1:2013ktq}%
  \BibitemOpen
  \bibfield  {author} {\bibinfo {author} {\bibfnamefont {V.}~\bibnamefont
  {Andreev}} \emph {et~al.} (\bibinfo {collaboration} {H1}),\ }\bibfield
  {title} {\bibinfo {title} {{Measurement of inclusive $e p$ cross sections at
  high $Q^2$ at $\sqrt s =$ 225 and 252 GeV and of the longitudinal proton
  structure function $F_L$ at HERA}},\ }\href
  {https://doi.org/10.1140/epjc/s10052-014-2814-6} {\bibfield  {journal}
  {\bibinfo  {journal} {Eur. Phys. J. C}\ }\textbf {\bibinfo {volume} {74}},\
  \bibinfo {pages} {2814} (\bibinfo {year} {2014})},\ \Eprint
  {https://arxiv.org/abs/1312.4821} {arXiv:1312.4821 [hep-ex]} \BibitemShut
  {NoStop}%
\bibitem [{\citenamefont {Wang}\ \emph {et~al.}(2022)\citenamefont {Wang},
  \citenamefont {Kou}, \citenamefont {Xie}, \citenamefont {Xie},\ and\
  \citenamefont {Chen}}]{Wang:2022jwh}%
  \BibitemOpen
  \bibfield  {author} {\bibinfo {author} {\bibfnamefont {X.}~\bibnamefont
  {Wang}}, \bibinfo {author} {\bibfnamefont {W.}~\bibnamefont {Kou}}, \bibinfo
  {author} {\bibfnamefont {G.}~\bibnamefont {Xie}}, \bibinfo {author}
  {\bibfnamefont {Y.-P.}\ \bibnamefont {Xie}},\ and\ \bibinfo {author}
  {\bibfnamefont {X.}~\bibnamefont {Chen}},\ }\bibfield  {title} {\bibinfo
  {title} {{Exclusive vector meson production with the analytical solution of
  Balitsky-Kovchegov equation}},\ }\href
  {https://doi.org/10.1088/1674-1137/ac6daa} {\bibfield  {journal} {\bibinfo
  {journal} {Chin. Phys. C}\ }\textbf {\bibinfo {volume} {46}},\ \bibinfo
  {pages} {093101} (\bibinfo {year} {2022})},\ \Eprint
  {https://arxiv.org/abs/2205.02396} {arXiv:2205.02396 [hep-ph]} \BibitemShut
  {NoStop}%
\bibitem [{\citenamefont {Chekanov}\ \emph {et~al.}(2004)\citenamefont
  {Chekanov} \emph {et~al.}}]{ZEUS:2004yeh}%
  \BibitemOpen
  \bibfield  {author} {\bibinfo {author} {\bibfnamefont {S.}~\bibnamefont
  {Chekanov}} \emph {et~al.} (\bibinfo {collaboration} {ZEUS}),\ }\bibfield
  {title} {\bibinfo {title} {{Exclusive electroproduction of J/psi mesons at
  HERA}},\ }\href {https://doi.org/10.1016/j.nuclphysb.2004.06.034} {\bibfield
  {journal} {\bibinfo  {journal} {Nucl. Phys. B}\ }\textbf {\bibinfo {volume}
  {695}},\ \bibinfo {pages} {3} (\bibinfo {year} {2004})},\ \Eprint
  {https://arxiv.org/abs/hep-ex/0404008} {arXiv:hep-ex/0404008} \BibitemShut
  {NoStop}%
\bibitem [{\citenamefont {Chekanov}\ \emph {et~al.}(2007)\citenamefont
  {Chekanov} \emph {et~al.}}]{ZEUS:2007iet}%
  \BibitemOpen
  \bibfield  {author} {\bibinfo {author} {\bibfnamefont {S.}~\bibnamefont
  {Chekanov}} \emph {et~al.} (\bibinfo {collaboration} {ZEUS}),\ }\bibfield
  {title} {\bibinfo {title} {{Exclusive rho0 production in deep inelastic
  scattering at HERA}},\ }\href {https://doi.org/10.1186/1754-0410-1-6}
  {\bibfield  {journal} {\bibinfo  {journal} {PMC Phys. A}\ }\textbf {\bibinfo
  {volume} {1}},\ \bibinfo {pages} {6} (\bibinfo {year} {2007})},\ \Eprint
  {https://arxiv.org/abs/0708.1478} {arXiv:0708.1478 [hep-ex]} \BibitemShut
  {NoStop}%
\bibitem [{\citenamefont {Adloff}\ \emph {et~al.}(2000)\citenamefont {Adloff}
  \emph {et~al.}}]{H1:1999pji}%
  \BibitemOpen
  \bibfield  {author} {\bibinfo {author} {\bibfnamefont {C.}~\bibnamefont
  {Adloff}} \emph {et~al.} (\bibinfo {collaboration} {H1}),\ }\bibfield
  {title} {\bibinfo {title} {{Elastic electroproduction of rho mesons at
  HERA}},\ }\href {https://doi.org/10.1007/s100520050703} {\bibfield  {journal}
  {\bibinfo  {journal} {Eur. Phys. J. C}\ }\textbf {\bibinfo {volume} {13}},\
  \bibinfo {pages} {371} (\bibinfo {year} {2000})},\ \Eprint
  {https://arxiv.org/abs/hep-ex/9902019} {arXiv:hep-ex/9902019} \BibitemShut
  {NoStop}%
\bibitem [{\citenamefont {Aktas}\ \emph {et~al.}(2006)\citenamefont {Aktas}
  \emph {et~al.}}]{H1:2005dtp}%
  \BibitemOpen
  \bibfield  {author} {\bibinfo {author} {\bibfnamefont {A.}~\bibnamefont
  {Aktas}} \emph {et~al.} (\bibinfo {collaboration} {H1}),\ }\bibfield  {title}
  {\bibinfo {title} {{Elastic J/psi production at HERA}},\ }\href
  {https://doi.org/10.1140/epjc/s2006-02519-5} {\bibfield  {journal} {\bibinfo
  {journal} {Eur. Phys. J. C}\ }\textbf {\bibinfo {volume} {46}},\ \bibinfo
  {pages} {585} (\bibinfo {year} {2006})},\ \Eprint
  {https://arxiv.org/abs/hep-ex/0510016} {arXiv:hep-ex/0510016} \BibitemShut
  {NoStop}%
\bibitem [{\citenamefont {Aaron}\ \emph {et~al.}(2010)\citenamefont {Aaron}
  \emph {et~al.}}]{H1:2009cml}%
  \BibitemOpen
  \bibfield  {author} {\bibinfo {author} {\bibfnamefont {F.~D.}\ \bibnamefont
  {Aaron}} \emph {et~al.} (\bibinfo {collaboration} {H1}),\ }\bibfield  {title}
  {\bibinfo {title} {{Diffractive Electroproduction of rho and phi Mesons at
  HERA}},\ }\href {https://doi.org/10.1007/JHEP05(2010)032} {\bibfield
  {journal} {\bibinfo  {journal} {JHEP}\ }\textbf {\bibinfo {volume} {05}},\
  \bibinfo {pages} {032}},\ \Eprint {https://arxiv.org/abs/0910.5831}
  {arXiv:0910.5831 [hep-ex]} \BibitemShut {NoStop}%
\bibitem [{\citenamefont {Golec-Biernat}\ and\ \citenamefont
  {Wusthoff}(1998)}]{Golec-Biernat:1998zce}%
  \BibitemOpen
  \bibfield  {author} {\bibinfo {author} {\bibfnamefont {K.~J.}\ \bibnamefont
  {Golec-Biernat}}\ and\ \bibinfo {author} {\bibfnamefont {M.}~\bibnamefont
  {Wusthoff}},\ }\bibfield  {title} {\bibinfo {title} {{Saturation effects in
  deep inelastic scattering at low Q**2 and its implications on diffraction}},\
  }\href {https://doi.org/10.1103/PhysRevD.59.014017} {\bibfield  {journal}
  {\bibinfo  {journal} {Phys. Rev. D}\ }\textbf {\bibinfo {volume} {59}},\
  \bibinfo {pages} {014017} (\bibinfo {year} {1998})},\ \Eprint
  {https://arxiv.org/abs/hep-ph/9807513} {arXiv:hep-ph/9807513} \BibitemShut
  {NoStop}%
\bibitem [{\citenamefont {Bartels}\ \emph {et~al.}(2002)\citenamefont
  {Bartels}, \citenamefont {Golec-Biernat},\ and\ \citenamefont
  {Kowalski}}]{Bartels:2002cj}%
  \BibitemOpen
  \bibfield  {author} {\bibinfo {author} {\bibfnamefont {J.}~\bibnamefont
  {Bartels}}, \bibinfo {author} {\bibfnamefont {K.~J.}\ \bibnamefont
  {Golec-Biernat}},\ and\ \bibinfo {author} {\bibfnamefont {H.}~\bibnamefont
  {Kowalski}},\ }\bibfield  {title} {\bibinfo {title} {{A modification of the
  saturation model: DGLAP evolution}},\ }\href
  {https://doi.org/10.1103/PhysRevD.66.014001} {\bibfield  {journal} {\bibinfo
  {journal} {Phys. Rev. D}\ }\textbf {\bibinfo {volume} {66}},\ \bibinfo
  {pages} {014001} (\bibinfo {year} {2002})},\ \Eprint
  {https://arxiv.org/abs/hep-ph/0203258} {arXiv:hep-ph/0203258} \BibitemShut
  {NoStop}%
\bibitem [{\citenamefont {Schrempp}(2005)}]{Schrempp:2005vc}%
  \BibitemOpen
  \bibfield  {author} {\bibinfo {author} {\bibfnamefont {F.}~\bibnamefont
  {Schrempp}},\ }\bibfield  {title} {\bibinfo {title} {{Instanton-induced
  processes: An Overview}},\ }in\ \href@noop {} {\emph {\bibinfo {booktitle}
  {{HERA and the LHC: A Workshop on the Implications of HERA for LHC Physics:
  CERN - DESY Workshop 2004/2005 (Midterm Meeting, CERN, 11-13 October 2004;
  Final Meeting, DESY, 17-21 January 2005)}}}}\ (\bibinfo {year} {2005})\ pp.\
  \bibinfo {pages} {3--16},\ \Eprint {https://arxiv.org/abs/hep-ph/0507160}
  {arXiv:hep-ph/0507160} \BibitemShut {NoStop}%
\bibitem [{\citenamefont {Martin}\ \emph {et~al.}(1996)\citenamefont {Martin},
  \citenamefont {Roberts},\ and\ \citenamefont {Stirling}}]{Martin:1996as}%
  \BibitemOpen
  \bibfield  {author} {\bibinfo {author} {\bibfnamefont {A.~D.}\ \bibnamefont
  {Martin}}, \bibinfo {author} {\bibfnamefont {R.~G.}\ \bibnamefont
  {Roberts}},\ and\ \bibinfo {author} {\bibfnamefont {W.~J.}\ \bibnamefont
  {Stirling}},\ }\bibfield  {title} {\bibinfo {title} {{Parton distributions: A
  Study of the new HERA data, alpha-s, the gluon and p anti-p jet
  production}},\ }\href {https://doi.org/10.1016/0370-2693(96)01031-3}
  {\bibfield  {journal} {\bibinfo  {journal} {Phys. Lett. B}\ }\textbf
  {\bibinfo {volume} {387}},\ \bibinfo {pages} {419} (\bibinfo {year}
  {1996})},\ \Eprint {https://arxiv.org/abs/hep-ph/9606345}
  {arXiv:hep-ph/9606345} \BibitemShut {NoStop}%
\bibitem [{\citenamefont {Golec-Biernat}\ and\ \citenamefont
  {Wusthoff}(1999)}]{Golec-Biernat:1999qor}%
  \BibitemOpen
  \bibfield  {author} {\bibinfo {author} {\bibfnamefont {K.~J.}\ \bibnamefont
  {Golec-Biernat}}\ and\ \bibinfo {author} {\bibfnamefont {M.}~\bibnamefont
  {Wusthoff}},\ }\bibfield  {title} {\bibinfo {title} {{Saturation in
  diffractive deep inelastic scattering}},\ }\href
  {https://doi.org/10.1103/PhysRevD.60.114023} {\bibfield  {journal} {\bibinfo
  {journal} {Phys. Rev. D}\ }\textbf {\bibinfo {volume} {60}},\ \bibinfo
  {pages} {114023} (\bibinfo {year} {1999})},\ \Eprint
  {https://arxiv.org/abs/hep-ph/9903358} {arXiv:hep-ph/9903358} \BibitemShut
  {NoStop}%
\bibitem [{\citenamefont {Marquet}\ \emph {et~al.}(2005)\citenamefont
  {Marquet}, \citenamefont {Peschanski},\ and\ \citenamefont
  {Soyez}}]{Marquet:2005ic}%
  \BibitemOpen
  \bibfield  {author} {\bibinfo {author} {\bibfnamefont {C.}~\bibnamefont
  {Marquet}}, \bibinfo {author} {\bibfnamefont {R.~B.}\ \bibnamefont
  {Peschanski}},\ and\ \bibinfo {author} {\bibfnamefont {G.}~\bibnamefont
  {Soyez}},\ }\bibfield  {title} {\bibinfo {title} {{QCD traveling waves at
  non-asymptotic energies}},\ }\href
  {https://doi.org/10.1016/j.physletb.2005.09.035} {\bibfield  {journal}
  {\bibinfo  {journal} {Phys. Lett. B}\ }\textbf {\bibinfo {volume} {628}},\
  \bibinfo {pages} {239} (\bibinfo {year} {2005})},\ \Eprint
  {https://arxiv.org/abs/hep-ph/0509074} {arXiv:hep-ph/0509074} \BibitemShut
  {NoStop}%
\bibitem [{\citenamefont {M\"antysaari}\ and\ \citenamefont
  {Schenke}(2018)}]{Mantysaari:2018zdd}%
  \BibitemOpen
  \bibfield  {author} {\bibinfo {author} {\bibfnamefont {H.}~\bibnamefont
  {M\"antysaari}}\ and\ \bibinfo {author} {\bibfnamefont {B.}~\bibnamefont
  {Schenke}},\ }\bibfield  {title} {\bibinfo {title} {{Confronting impact
  parameter dependent JIMWLK evolution with HERA data}},\ }\href
  {https://doi.org/10.1103/PhysRevD.98.034013} {\bibfield  {journal} {\bibinfo
  {journal} {Phys. Rev. D}\ }\textbf {\bibinfo {volume} {98}},\ \bibinfo
  {pages} {034013} (\bibinfo {year} {2018})},\ \Eprint
  {https://arxiv.org/abs/1806.06783} {arXiv:1806.06783 [hep-ph]} \BibitemShut
  {NoStop}%
\bibitem [{\citenamefont {Accardi}\ \emph {et~al.}(2016)\citenamefont {Accardi}
  \emph {et~al.}}]{Accardi:2012qut}%
  \BibitemOpen
  \bibfield  {author} {\bibinfo {author} {\bibfnamefont {A.}~\bibnamefont
  {Accardi}} \emph {et~al.},\ }\bibfield  {title} {\bibinfo {title} {{Electron
  Ion Collider: The Next QCD Frontier}: {Understanding the glue that binds us
  all}},\ }\href {https://doi.org/10.1140/epja/i2016-16268-9} {\bibfield
  {journal} {\bibinfo  {journal} {Eur. Phys. J. A}\ }\textbf {\bibinfo {volume}
  {52}},\ \bibinfo {pages} {268} (\bibinfo {year} {2016})},\ \Eprint
  {https://arxiv.org/abs/1212.1701} {arXiv:1212.1701 [nucl-ex]} \BibitemShut
  {NoStop}%
\bibitem [{\citenamefont {Abdul~Khalek}\ \emph {et~al.}(2021)\citenamefont
  {Abdul~Khalek} \emph {et~al.}}]{AbdulKhalek:2021gbh}%
  \BibitemOpen
  \bibfield  {author} {\bibinfo {author} {\bibfnamefont {R.}~\bibnamefont
  {Abdul~Khalek}} \emph {et~al.},\ }\bibfield  {title} {\bibinfo {title}
  {{Science Requirements and Detector Concepts for the Electron-Ion Collider:
  EIC Yellow Report}},\ }\href@noop {} {\  (\bibinfo {year} {2021})},\ \Eprint
  {https://arxiv.org/abs/2103.05419} {arXiv:2103.05419 [physics.ins-det]}
  \BibitemShut {NoStop}%
\bibitem [{\citenamefont {Chen}(2018)}]{Chen:2018wyz}%
  \BibitemOpen
  \bibfield  {author} {\bibinfo {author} {\bibfnamefont {X.}~\bibnamefont
  {Chen}},\ }\bibfield  {title} {\bibinfo {title} {{A Plan for Electron Ion
  Collider in China}},\ }\href {https://doi.org/10.22323/1.316.0170} {\bibfield
   {journal} {\bibinfo  {journal} {PoS}\ }\textbf {\bibinfo {volume}
  {DIS2018}},\ \bibinfo {pages} {170} (\bibinfo {year} {2018})},\ \Eprint
  {https://arxiv.org/abs/1809.00448} {arXiv:1809.00448 [nucl-ex]} \BibitemShut
  {NoStop}%
\bibitem [{\citenamefont {Chen}\ \emph {et~al.}(2020)\citenamefont {Chen},
  \citenamefont {Guo}, \citenamefont {Roberts},\ and\ \citenamefont
  {Wang}}]{Chen:2020ijn}%
  \BibitemOpen
  \bibfield  {author} {\bibinfo {author} {\bibfnamefont {X.}~\bibnamefont
  {Chen}}, \bibinfo {author} {\bibfnamefont {F.-K.}\ \bibnamefont {Guo}},
  \bibinfo {author} {\bibfnamefont {C.~D.}\ \bibnamefont {Roberts}},\ and\
  \bibinfo {author} {\bibfnamefont {R.}~\bibnamefont {Wang}},\ }\bibfield
  {title} {\bibinfo {title} {{Selected Science Opportunities for the EicC}},\
  }\href {https://doi.org/10.1007/s00601-020-01574-0} {\bibfield  {journal}
  {\bibinfo  {journal} {Few Body Syst.}\ }\textbf {\bibinfo {volume} {61}},\
  \bibinfo {pages} {43} (\bibinfo {year} {2020})},\ \Eprint
  {https://arxiv.org/abs/2008.00102} {arXiv:2008.00102 [hep-ph]} \BibitemShut
  {NoStop}%
\bibitem [{\citenamefont {Anderle}\ \emph {et~al.}(2021)\citenamefont {Anderle}
  \emph {et~al.}}]{Anderle:2021wcy}%
  \BibitemOpen
  \bibfield  {author} {\bibinfo {author} {\bibfnamefont {D.~P.}\ \bibnamefont
  {Anderle}} \emph {et~al.},\ }\bibfield  {title} {\bibinfo {title}
  {{Electron-ion collider in China}},\ }\href
  {https://doi.org/10.1007/s11467-021-1062-0} {\bibfield  {journal} {\bibinfo
  {journal} {Front. Phys. (Beijing)}\ }\textbf {\bibinfo {volume} {16}},\
  \bibinfo {pages} {64701} (\bibinfo {year} {2021})},\ \Eprint
  {https://arxiv.org/abs/2102.09222} {arXiv:2102.09222 [nucl-ex]} \BibitemShut
  {NoStop}%
\end{thebibliography}%

\end{document}